\documentclass[aps, prfluids, preprint, superscriptaddress]{revtex4-1}

\usepackage{epsfig}
\usepackage{graphicx}
\usepackage{amssymb}
\usepackage{amsmath}
\usepackage{comment}
\usepackage{textcomp} 
\usepackage{color, soul}
\usepackage{natbib}
\usepackage{mathtools}
\usepackage{gensymb}
\usepackage{hyperref}
\usepackage{graphicx}
\usepackage{subcaption}
\usepackage{epstopdf}
\usepackage{comment}
\usepackage{adjustbox}
\usepackage{booktabs}
\usepackage{array}
\usepackage{multirow} % For merging rows
\usepackage{float}
\usepackage{mathtools}
\usepackage{tabularx}
\usepackage{graphicx}
\usepackage{xcolor}
\definecolor{FuchsiaPink}{rgb}{0.78, 0.08, 0.52}
\usepackage{pgfplots}
\usepackage{pgfplotstable}
\usepgfplotslibrary{fillbetween}
\usetikzlibrary{shapes.geometric}
\pgfplotsset{
  log x ticks with fixed point/.style={
      xticklabel={
        \pgfkeys{/pgf/fpu=true}
        \pgfmathparse{exp(\tick)}%
        \pgfmathprintnumber[fixed relative, precision=3]{\pgfmathresult}
        \pgfkeys{/pgf/fpu=false}
      }
  },
  log y ticks with fixed point/.style={
      yticklabel={
        \pgfkeys{/pgf/fpu=true}
        \pgfmathparse{exp(\tick)}%
        \pgfmathprintnumber[fixed relative, precision=3]{\pgfmathresult}
        \pgfkeys{/pgf/fpu=false}
      }
  }
}
\pgfplotsset{compat=1.3}
\usepackage{tikz}
\usepackage{pgfplots}
\pgfplotsset{width=10cm,compat=1.9}
\usepgfplotslibrary{external}
\usetikzlibrary{external}
\usepackage{pgfplotstable}
\usepackage{csvsimple}
\usepgfplotslibrary{fillbetween}
\pgfplotsset{compat=1.9}
\tikzset{external/system call={pdflatex \tikzexternalcheckshellescape -halt-on-error
-interaction=batchmode -jobname "\image" "\texsource"}}

\usepackage[capitalize]{cleveref}

\newcommand{\strkvol}{\mathop{\ooalign{\hfil$V$\hfil\cr\kern0.08em--\hfil\cr}}\nolimits}

\begin{document}

\title{Entry and penetration of a superhydrophobic sphere into a deep pool}
 \author{Prasanna Kumar Billa}
\affiliation{Multiscale Multiphysics Group (MMG), Department of Mechanical Engineering, Indian Institute of Technology Madras, 600036, India}

 \author{Cameron Tropea}
 \email{tropea@sla.tu-darmstadt.de}
 \affiliation{Multiscale Multiphysics Group (MMG), Department of Mechanical Engineering, Indian Institute of Technology Madras, 600036, India}
\affiliation{Institute for Fluid Mechanics and Aerodynamics, Technical University of Darmstadt, 64287, Germany}

 \author{Pallab Sinha Mahapatra}
\email{pallab@iitm.ac.in}
\affiliation{Multiscale Multiphysics Group (MMG), Department of Mechanical Engineering, Indian Institute of Technology Madras, 600036, India}

\begin{abstract}
This study experimentally examines the entry and penetration of a superhydrophobic sphere into a quiescent deep pool, with special emphasis placed on the primary and secondary pinch-off of the air cavity existing in its wake. Two aspects are novel in this study. For one, the experiments are performed for a large range of dimensionless sphere densities, where lighter spheres, with their air cavity, exhibit a terminally ascending trajectory and heavier spheres a terminally descending trajectory. The second novel result is a strong correlation of primary and secondary pinch-off times with the  Froude number at impact and the dimensionless density. A semi-empirical correlation for the air cavity volume following the primary pinch-off shows excellent agreement with measurements over all dimensionless densities. A scalar force balance predicts a drastic decrease of buoyancy upon pinch-off, reflected also in the abrupt change of deceleration, measured using two orthogonally placed high-speed cameras to capture the time resolved trajectory of the sphere in the pool. Comparisons are drawn between the trajectories of superhydrophobic spheres and those of hydrophilic spheres, measured in a previous study.

\end{abstract}


\maketitle
\section{Introduction}

The entry and penetration of solid objects into a deep quiescent liquid pool occurs in numerous applications, ranging from industrial processes, naval engineering, military, sports, as well as many natural events. The case of a sphere entering the pool has often been used to generically represent this flow situation, whereby the sphere wettability plays a decisive role throughout both the entry and the penetration stages. Hydrophilic spheres can enter the pool with no perceptible air cavity in the wake, depending on the impact Reynolds number \cite{kuwabara1983anomalous, horowitz2010effect,Brown2003SphereRevisited}. On the other hand, hydrophobic spheres result in an air cavity, which exhibits a pinch-off during the first phase of penetration, leaving a residual air cavity in the wake of the sphere \cite{Duclaux2007DynamicsCavities,Aristoff2009WaterSpheres,Truscott2012UnsteadyEntry,Vakarelski2017Self-determinedCavities}. The size and persistence of the wake cavity depends on numerous factors, in particular, the impact Reynolds number, density of the sphere, the degree of hydrophobicity, impact angle \cite{asfar1987rigid}, possible sphere rotation \cite{truscott2009cavity, techet2011water}, etc. The present study examines the case of such superhydrophobic spheres.

There have been a large number of past studies on the entry of hydrophobic spheres into deep pools in which attention has been directed toward various features and/or stages of the process. For instance, some have examined the splash condition and characteristics upon impact \cite{may1951effect,marston2016crown,watson2020making,sun2019splash}, whereas others have concentrated on the hydrodynamics of the cavity formation and its pinch-off, and still others are more concerned with the trajectory and penetration of the sphere after the submersion phase.

In a seminal study, \citet{Aristoff2009WaterSpheres} examined the cavity types formed by the impact of superhydrophobic spheres across a wide range of Weber and Bond numbers, specifically focusing on high density steel spheres and steel cylinders. Theoretical models were developed to describe the shape of the air cavity in the investigated Bond number regime, based on the Besant-Rayleigh problem. This work identified different cavity shapes, classified as quasi-static seal, shallow seal, deep seal, and surface seal, depending on the length of the cavity. In the present study, within our Weber number limits, we concentrate on the deep seal cavity formation. \citet{Aristoff2010TheSpheres} further advanced the understanding of pinch-off dynamics by proposing a theoretical model that predicts key pinch-off parameters, including pinch-off time, cavity length, total cavity volume, and air entrainment, for spheres of various densities. Building on this, \citet{Truscott2012UnsteadyEntry} developed a force model to estimate the unsteady forces acting on both hydrophilic and superhydrophobic spheres during water entry. Their study, supported by Particle Image Velocimetry (PIV), confirmed the formation of vortices in the wake of the spheres.

Full characterization of the cavity dynamics of a surface wettability modified glass sphere is presented experimentally and analytically in \citet{Duclaux2007DynamicsCavities}. In this study, an approximate analytical model based on the classical Besant-Rayleigh problem was utilized to investigate the temporal radial expansion and collapse time of the cavity following the impact of the sphere from the air. Later, the effect of surface tension and fluid viscosity on the normal impact of hydrophobic spheres on the water surface is reported for low Bond numbers \cite{aristoff2008water,Aristoff2009WaterSpheres}.

The drag reduction of superhydrophobic spheres and fully cavity encapsulated spheres has been studied for various impacting conditions \cite{mchale2009terminal, vakarelski2011drag, Vakarelski2017Self-determinedCavities, jetly2018drag}. For instance, \citet{li2019experimental} investigated a sphere with high density ($7896$ $\mathrm{kg/m^3}$) and varying surface wettability. Their study focused on the splash evolution above the interface, cavity dynamics, and hydrodynamics of these spheres at different impact velocities. They concluded that, consistent with findings from \cite{Truscott2012UnsteadyEntry}, the hydrodynamic force coefficient derived from instantaneous velocities and accelerations is higher for the hydrophilic sphere than the superhydrophobic sphere.

 Other studies were more directed toward the sphere penetration phase, after the primary pinch-off of the entry cavity. For instance, 
\citet{Mansoor2017Stable-streamlinedSpheres} quantitatively examined the drag forces and hydrodynamic behaviour of superhydrophobic and Leidenfrost spheres, which generate stable, streamlined, and unstable cavities upon impact, providing detailed force measurements. The hydrodynamic forces acting on the sphere were evaluated by considering the contributions of inertia, added mass, gravitation, buoyancy, and drag forces. 

Studies using hydrophobic spheres consistently demonstrate reduced hydrodynamic forces compared to hydrophilic spheres, a conclusion that aligns with earlier findings \cite{Truscott2012UnsteadyEntry,li2019experimental}. Extensive experimental and numerical investigations have been carried out to further understand these phenomena, including those by \citet{vakarelski2011drag,Truscott2012UnsteadyEntry,Mansoor2014WaterFormation,Mansoor2017Stable-streamlinedSpheres,Vakarelski2014LeidenfrostWater,Vakarelski2017Self-determinedCavities,li2019experimental,Guleria2021ExperimentalSphere}. Despite these advances, the influence of the closed cavity following the pinch-off event on the subsequent dynamics of the sphere has remained relatively unexplored.

In summary, previous studies have focused largely on the impact dynamics, air cavity formation, and pinch-off behavior of rigid superhydrophobic spheres with high density ratios (sphere-to-water), leaving a significant lack of investigations for spheres of lower densities ($<$  3000 $ \mathrm{kg/m^3}  $ ), particularly regarding their trajectory and force evolution during descent over extended time scales. The present study aims to bridge this gap by systematically investigating a wide range of sphere densities, diameters, and impact velocities, comparing the behavior of superhydrophobic spheres to their hydrophilic counterparts under identical conditions by evaluating the time resolved forces using high-speed trajectory imaging. Further, we show that the buoyancy term undergoes strong and abrupt jumps in magnitude, associated with the primary and any secondary pinch-off events, ultimately deciding whether the sphere will be terminally descending or will start to ascend back to the free surface. In addition, the drag coefficient undergoes significant changes with pinch-off events, generally exhibiting lower values than for hydrophilic spheres, which have no air cavity in their wake.

\section{\label{sec:experimental details}Experimental set-up and methodology}
 
\subsection{Experimental set-up}
\begin{figure}[t]
\centering
\includegraphics[width=0.8\textwidth]{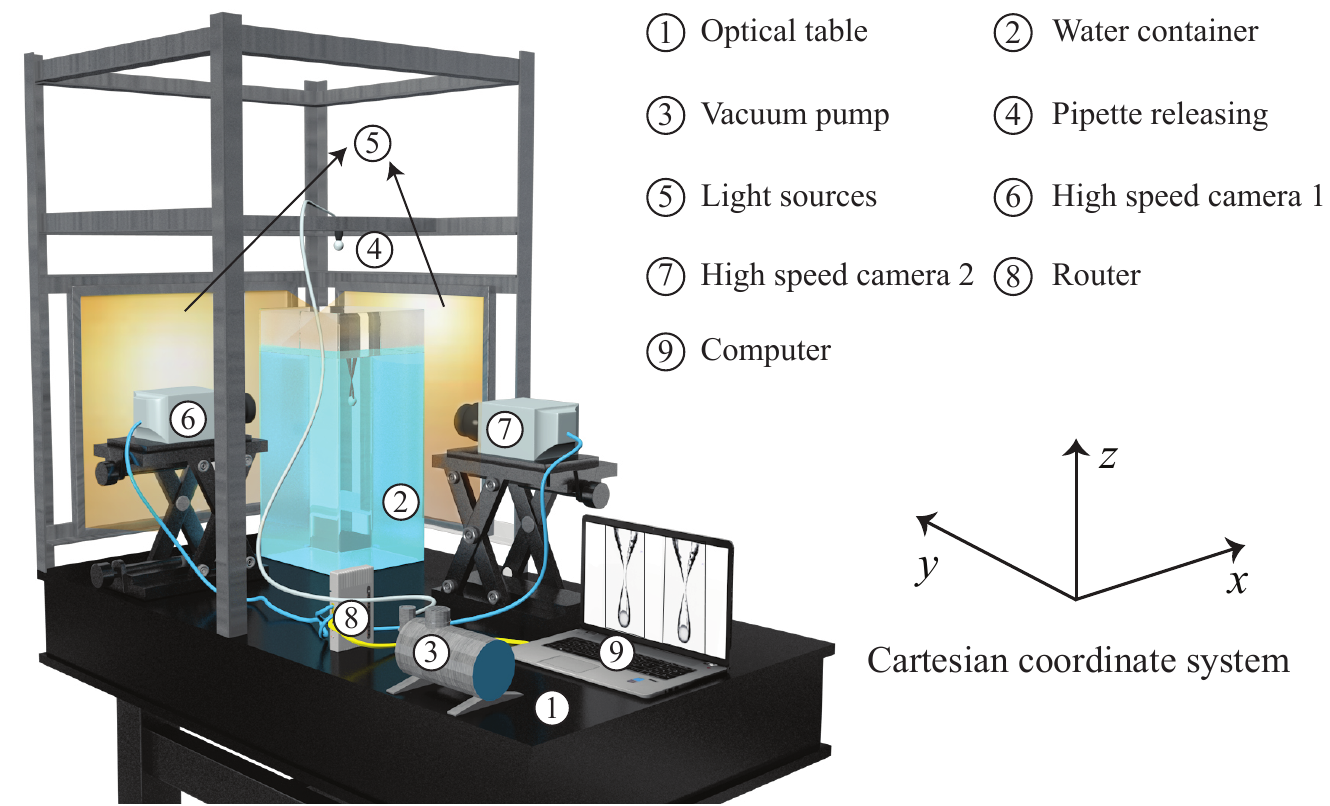}
\caption{(color online) Schematic of the experimental set-up. The container has dimensions 200(L) $\times$ 200(B) $\times$ 400(H)$ \mathrm{mm}^3$, whereby the largest sphere used in the study was 10~mm. The impact velocity of the spheres is varied by varying the height of sphere release.} 
\label{fig:two_cameras}
\end{figure}

The experimental set-up is identical to that used for a previous study devoted to the impact and penetration of hydrophilic spheres \citep{Billa_Josyula_Tropea_Mahapatra_2025} and is only briefly described here. The set-up is pictured in Fig.~\ref{fig:two_cameras}, comprising a clear, translucent acrylic container, two orthogonally placed, synchronized high-speed monochrome cameras (Phantom VEO E-340-L with a Nikon microlens 28 mm, spatial resolution of 192 $\mu$m/pixel and Phantom VEO E-340-L with a Nikon micro lens of focal length 24-85 mm, spatial resolution of 199 $\mu$m/pixel), two back illumination panels (30W) and a vacuum system used to release the spheres using a pipette. The cameras  operate at 1000~fps with a resolution of 1152 $\times$ 1100 pixels and an exposure time of 400~$\mu$s. The field of view in both the cameras is 100~mm $\times$ 160 mm. 

\subsection{Preparation of superhydrophobic spheres}
\label{SH-preparation}
Spheres were made superhydrophobic (SH)  using a commercially available coating (NeverWet spray). The spray is a two-step system, designed to create a moisture repelling barrier on a variety of substrates.  First, the wettable spheres are cleaned with alcohol and ethanol as described in \cite{truscott2009cavity}. The cleaned spheres are placed in a dust free petri dish for 20~min and are arranged so that they do not come into contact with each other or the wall. The base coat is sprayed twice and the spheres are then cured at ambient temperature for 30~min before applying the top coat. The spheres are then placed in a dry environment for 12 hours before performing the experiment. 

 The static contact angle on the spheres, measured using the ellipse fit method in ImageJ and accounting for sphere curvature \cite{Extrand2002WaterSurfaces, Extrand2012IndirectSurfaces, Magos2021ContactSurfaces}, was 155 $\pm$ 5$^\circ$. Although the low hysteresis makes this measurement challenging, \citet{gupta2016superhydrophobic} and \citet{weisensee2016water} also measured the contact angle on flat substrates (glass: 2700 kg/m\textsuperscript{3}, polymethylmethacrylate: 1180 kg/m\textsuperscript{3}) coated with NeverWet spray, reporting advancing and receding angles of 164 $\pm$ 4$^\circ$ and 159 $\pm$ 3$^\circ$, respectively. This confirms that the NeverWet spray renders the surface extremely water repellent. 

\subsection{Range of varied parameters}
\begin{table}[b]
\caption{\label{tab:parameters}
Definitions and range of varied parameters
}
\begin{ruledtabular}
\begin{tabular}{{c c c c }}
\textrm{Parameter}&
\textrm{Symbol}&
\textrm{Definition}&
\textrm{Range of values}\\
\colrule
        Sphere diameter & $D$ & -  & 4, 6, 10~mm  \\
       Density ratio &$\rho^*$ & $\rho_s/\rho_l$  & 1.12, 1.37, 2.16, 3.26, 6.08, 7.92  \\
       Impact velocity & $v_i$ & $ \sqrt{2gh} $  & 0.89 - 3.52~m/s \\
       Impact Reynolds number & $\mathrm{Re}_i$ & $v_iD/\nu$  & 4,000 - 39,600  \\
     Impact Froude number & $\mathrm{Fr}_i$ & $ v_i/\sqrt{gD}$  & 3.07 - 17.32  \\
\end{tabular}
\end{ruledtabular}
\end{table}

Table~\ref{tab:parameters} provides the parameters and dimensionless quantities pertinent to this study. The diameter ($D$) of the spheres is measured using a Vernier caliper and their mass ($m$) is determined using an electronic weight balance (Ohaus), yielding the sphere density $\rho_s$. The density of the liquid $\rho_l$ is taken as 1000 $\mathrm{kg/m^3}$. The impact velocity ($v_i \approx \sqrt{ 2gh }$) is derived from the initial release height ($h$) of the sphere, measured from its center to the air-water interface. The operational ranges of impact Reynolds and Froude numbers are provided in Table~\ref{tab:parameters}. In these definitions $\nu$ is the kinematic viscosity (0.89 $ \mathrm{mm^2/s}$) of the fluid and $g$ is gravitational acceleration (9.81~$\mathrm{m/s^2}$) \cite{pallas1983surface}.
Lengths are rendered dimensionless using the sphere diameter $D$, velocities using $v_i$, and time using $D/v_i$. Dimensionless quantities are designated with a * superscript.

One distinguishing feature of the entry of superhydrophic spheres is the formation of an air cavity in its wake, which, after some penetration length/time pinches off, leaving an attached wake air cavity, as shown in Fig.~\ref{fig:Displaced_Volume}. Hydrophilic spheres exhibit no similar air cavity up to some limiting impact Reynolds number, since the contact line upon entry moves as rapidly as the sphere penetration, completely wetting the sphere up to its apex (\cite{Duclaux2007DynamicsCavities}, Fig.~5(a)). Under what conditions an air cavity forms has been the subject of numerous investigations and modeling, starting with \cite{may1951effect} and further pursued by \cite{duez2007making, Duclaux2007DynamicsCavities} and \cite{Aristoff2009WaterSpheres}. \citet{Truscott2012UnsteadyEntry} examined the entry of hydrophobic spheres and modeled the cavity growth and collapse using a potential flow model, as in \cite{Duclaux2007DynamicsCavities}. \citet{Speirs2019TheJet} also characterized the influence of contact angle on the water entry of a sphere. Nevertheless, few studies exist using superhydrophobic spheres (maximum advancing contact angle of \citet{Speirs2019WaterAnglesb} was 141$^\circ$). Due to the large contact angle ($> 150^\circ $), all spheres in the present study exhibited an air cavity upon impact and a pinch-off, resulting in a residual air cavity in their wake, regardless of their impact velocity \cite{Truscott2014WaterProjectiles}.

\subsection*{Image processing}

\begin{figure}
\centering
\includegraphics[width=0.6\textwidth]{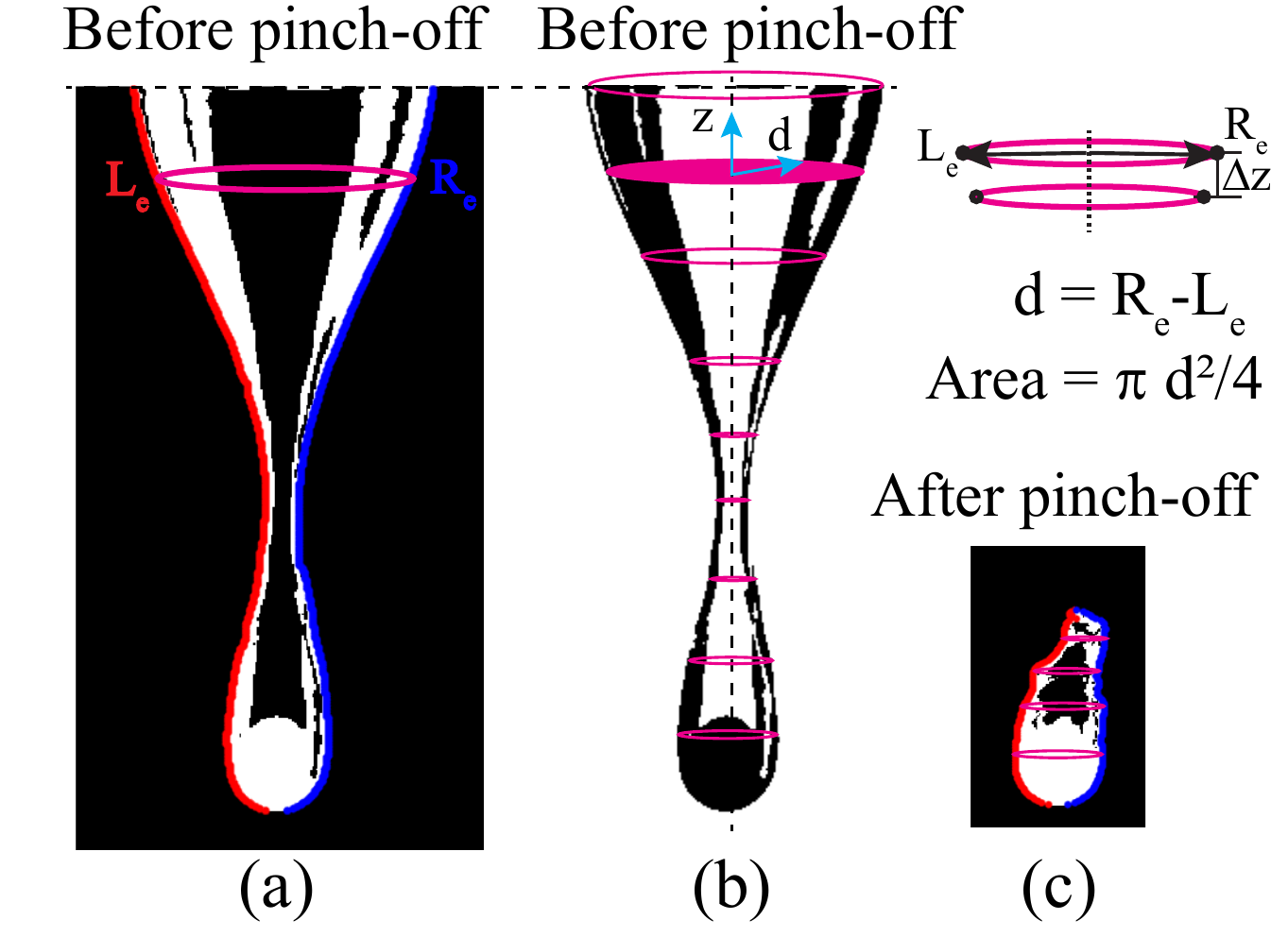}
\caption{ (color online) Computation of displaced volume. (a) Each image sequence is converted into a binary image, with the left and right edges detected to define the local diameter. (b) The cross-sectional diameter $d$ at each vertical position is used to find the area of a disk. (c) The displaced volume is obtained by integrating over the vertical extent of the cavity using a disk thickness of $\Delta z$, with the integration performed using the trapezoidal method.}
\label{fig:Displaced_Volume}
\end{figure}

 The buoyancy force acting on the completely submerged spheres is computed using the displaced volume of fluid ($V_\mathrm{Disp}$) as  $F_\mathrm{B}=\rho_lgV_\mathrm{Disp}$ (Archimedes principle), thus, the combined displaced volume of the sphere and the entrained air must be determined for each image frame. The displaced volume before and after pinch-off is determined according to the process pictured in Fig.~\ref{fig:Displaced_Volume} and begins with background subtraction and conversion to binary format. A Canny edge detection algorithm in MATLAB is applied to identify the left ($L_e$) and right ($R_e$) boundaries of the cavity and the sphere, as shown in Fig.~\ref{fig:Displaced_Volume}(a). The diameter at each vertical position is computed as \( d=R_e - L_e \) and used to determine the cross-sectional area of each horizontal disk (see Fig.~\ref{fig:Displaced_Volume}(b)). 
To estimate the total displaced volume ($V_\mathrm{Disp}$), the disk areas are integrated along the vertical axis using the trapezoidal rule with a disk height of $\Delta z$. In some experiments,  secondary pinch-offs of the attached air cavity are observed, in which case the above process is again applied to the remaining cavity-sphere displaced volume (see Fig.~\ref{fig:Displaced_Volume}(c)).  
The displaced volume is rendered dimensionless with the volume of the sphere ($V_\mathrm{S}$). Accordingly, the dimensionless displaced volume is defined as $V_\mathrm{Disp}^* = V_\mathrm{Disp}/V_\mathrm{S}$. The above procedure implicitly assumes that the cavity is axisymmetric.

The accumulated dimensionless path traversed by the sphere is represented as $s^* = \sqrt{x^2+y^2+z^2}/D$. Although the trajectory of the spheres exhibited very little deviation from a pure vertical motion, a dimensionless lateral displacement was computed for each time step using $r^*=\sqrt{x^2 + y^2}/D$. The velocity magnitude $|\Vec{v}_s | $ is computed by differentiating the displacement data after first fitting that data with a quintic spline function. 
The acceleration ($a_s^*$) is then computed from the time derivative of the velocity. To reduce noise in the acceleration ($a_s^*$), the velocity is first smoothed using a piecewise spline fit applied over fixed time intervals.  The resulting acceleration ($a_s^*$) is further smoothed using a smoothing spline, with a smoothing parameter close to one.

\subsection{Equation of motion}
\label{subsec:eos}
\begin{figure}[h]
   \begin{subfigure}{0.37\textwidth}
  	\includegraphics[width=\textwidth]{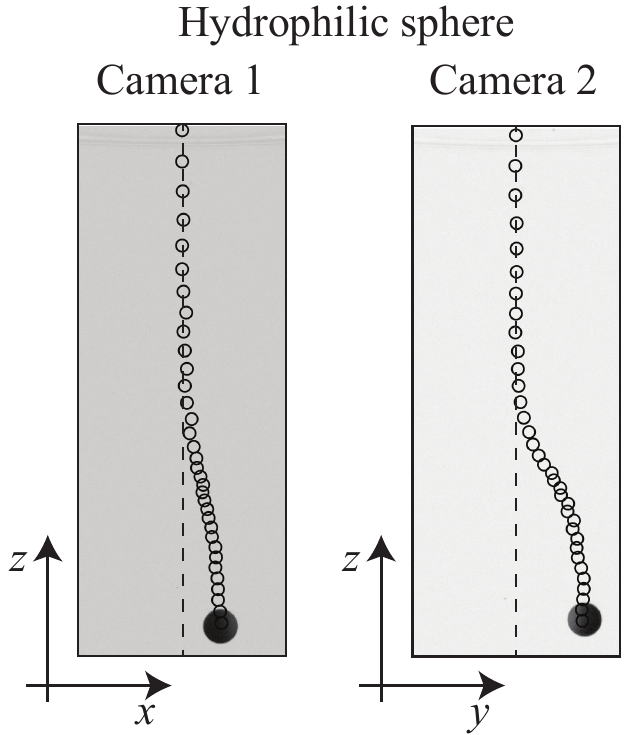}
	 \caption{}
   \end{subfigure}
   \hspace{0.1\textwidth}
   \begin{subfigure}{0.37\textwidth}
    		\includegraphics[width=\textwidth]{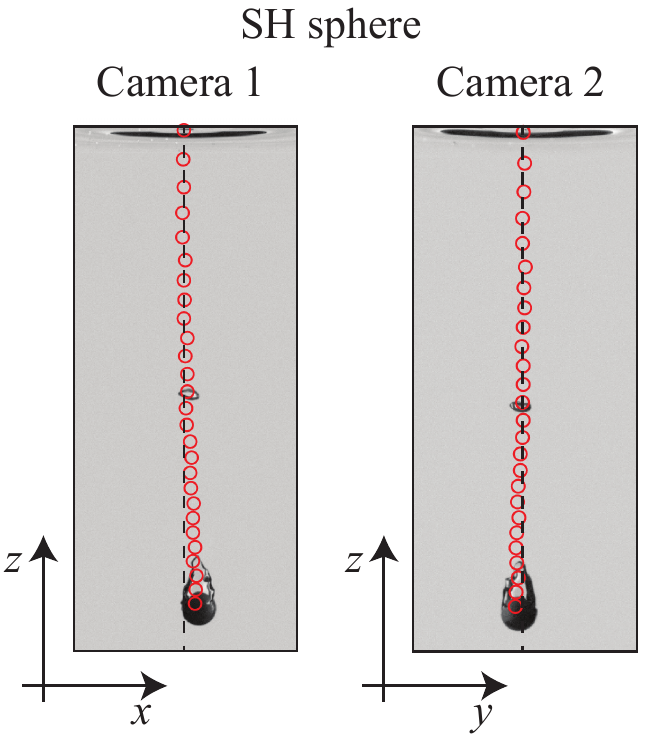}
   \caption{}
   \end{subfigure}
   %\centering
   \begin{subfigure}{0.37\textwidth}
    \includegraphics[width=\textwidth]{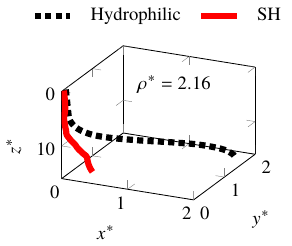}
    \caption{}
   \end{subfigure}
   \hspace{0.1\textwidth}
    \begin{subfigure}{0.37\textwidth}
    \centering
    \includegraphics[width=\textwidth]{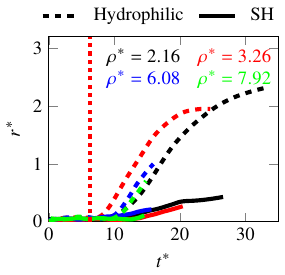}
         \caption{}
    \end{subfigure}
    \caption{ (color online) The trajectories of a 10~mm diameter hydrophilic and SH sphere with a density ratio $\rho^*=2.16$ at Re$_i=15,700$. The trajectories viewed from the orthogonally placed high-speed cameras for the (a)  hydrophilic sphere and (b) SH sphere. (c) Three-dimensional rendition of the trajectories. (d) Comparison of the dimensionless lateral displacement ($r^*$) for 10~mm spheres of varying density at Re$_i$ = 15,700. The red dashed vertical line in (d) represents the dimensionless primary pinch-off time. }
    \label{fig:Result_NC10mm_Re15700}
\end{figure}

In contrast to previous measurements performed using hydrophilic spheres \cite{Billa_Josyula_Tropea_Mahapatra_2025}, the superhydrophobic spheres exhibited only very low values of $r^*$, suggesting that only weak lift/lateral forces were acting during its descent. This fact is illustrated with comparative trajectory results shown in Fig.~\ref{fig:Result_NC10mm_Re15700}(a), (b) and (c). In this figure, the trajectory traces of a 10~mm, $\rho^*=2.16$ superhydrophobic sphere are compared with those of a 10~mm hydrophilic sphere, both impacting with a Reynolds number of 15,700. A comparison of the lateral displacements for spheres of different density ratios is shown in Fig.~\ref{fig:Result_NC10mm_Re15700}(d). Given this apparent lack of lateral lift forces, the assumption can be made that the flow around the sphere, including the wake air cavity, is comparably more axialsymmetric. The unsteady and non-symmetric wake as a source of lift forces acting on hydrophilic spheres has been postulated by numerous authors in the past \cite{scoggins1967sphere,taneda1978visual,kuwabara1983anomalous,Veldhuis2005MotionParticles,veldhuis2007experimental} and in some cases confirmed by PIV measurements in the wake region \cite{horowitz2008critical,horowitz2010effect,Truscott2012UnsteadyEntry,Vakarelski2017Self-determinedCavities,jetly2018drag}. However, the symmetric wake behaviour with superhydrophobic spheres is in agreement with many other studies of superhydrophobic spheres penetrating a deep pool \cite{Aristoff2010TheSpheres,aristoff2008water,Truscott2012UnsteadyEntry,Speirs2019WaterAnglesb} and allows the implicit conclusion that the axial symmetry of the three-phase contact line arising with the superhydrophobic sphere is steadier than the boundary-layer separation around a hydrophilic sphere.

The lack of lift forces justifies the use of force terms in the equation of motion acting only in the $z$ direction, implicitly assuming that the $z$ direction is equivalent to the pathline (in-line). This scalar momentum equation (Boussinesq-Basset-Oseen, BBO) for the completely submerged sphere/air-cavity  then reads  \citep{zhu1998multiphase,crowe2011multiphase}:
\begin{eqnarray}
      \underbrace{\frac{1}{6} \rho_s \pi D^3 \frac{\mathrm{d}v_s}{\mathrm{d}t}}_\text{Inertial force}=  
      \underbrace{\frac{1}{6} \pi D^3\rho_sg}_\text{Gravity}-
      F_\textrm{B} 
        - F_\textrm{H}
        - F_\gamma
      - \underbrace{\frac{1}{6} C_A \rho_l \pi D^3 \frac{\mathrm{d}v_s}{\mathrm{d}t}}_\text{Added mass} \nonumber \\ 
      - \underbrace{\frac{3}{2} D ^2 \sqrt{\pi \mu \rho_l} \int_{0}^{t} \frac{1}{\sqrt{t-\zeta}} \frac{\mathrm{d}v_s}{\mathrm{d}\zeta}\mathrm{d}\zeta }_\text{Boussinesq-Basset term}, 
      \label{eq:Force_evaluation} 
\end{eqnarray}

\noindent where $F_\textrm{B}$, $F_\mathrm{H}$, and $F_\gamma$ are the buoyancy force, the hydrodynamic force, and forces arising from surface tension, respectively. $C_A$ is the coefficient of the added mass force. In this equation, the mass of the wake air cavity has been neglected in the inertial and gravity terms. 

Furthermore, no rotational forces have been considered, since observations with marked spheres confirmed that the spheres were not rotating. Normally, $F_\mathrm{H}$ would be interpreted as a drag force, expressed using a drag coefficient $C_\mathrm{D}$ (e.g. $1/8*C_D\rho_l \pi D^2 v_{s}^2$), but in this particular case further hydrodynamic forces arise (as discussed below in Sect.~\ref{sec:force balance}). In any case, in the expectation that $F_\textrm{H}$ will eventually be a drag force, it is assumed to act in the upward direction, i.e. it is subtracted from the right-hand side of the equation.

\begin{figure}
    \centering
    \begin{subfigure}{0.389\textwidth}
        \centering
        \includegraphics[width=\textwidth]{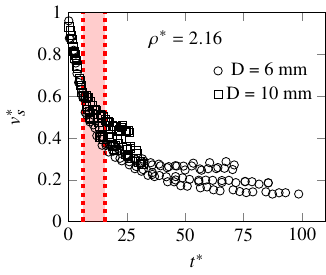}
        \caption{}
    \end{subfigure}
    \hspace{0.05\textwidth}
    \begin{subfigure}{0.4\textwidth}
        \centering
        \includegraphics[width=\textwidth]{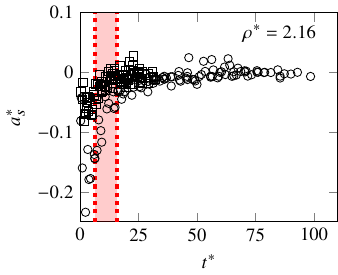}
        \caption{}
    \end{subfigure}
    \begin{subfigure}{0.389\textwidth}
        \centering
        \includegraphics[width=\textwidth]{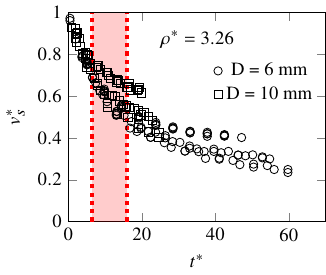}
        \caption{}
    \end{subfigure}
    \hspace{0.05\textwidth}
    \begin{subfigure}{0.4\textwidth}
        \centering
        \includegraphics[width=\textwidth]{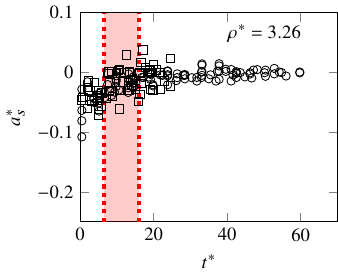}
        \caption{}
    \end{subfigure}
    \caption{(color online) Magnitude of dimensionless velocity ($v_s^*$) and acceleration ($a_s^*$) for  superhydrophobic spheres of diameter 6~mm and 10~mm and for all impact Reynolds numbers ($\mathrm{Re}_i$): (a, b) $\rho^* = 2.16$; (c, d) $\rho^* = 3.26$. The shaded region represents the dimensionless cavity pinch-off time under various impact conditions.}
    \label{fig:kinematics_all}
\end{figure}

We can expect that neither the added mass term (with a constant coefficient $C_\mathrm{A}$) nor the Basset history term in the form given in Eq.~(\ref{eq:Force_evaluation}) will be applicable to the decelerating sphere with an attached, fluctuating wake cavity. Even the drag coefficient is not expected to be constant in this decelerating state of motion, as measured in \cite{Billa_Josyula_Tropea_Mahapatra_2025}.  Building on work in \cite{dong1978effective}, \citet{Mansoor2014WaterFormation} used a constant added mass term of $F_\mathrm{A}=0.672 (1/6)\rho_l\pi D^3 (\mathrm{d}v_s/\mathrm{d}t)$ and did not consider the history term. These assumptions are discussed further below. 

In anticipation of further results presented below, Fig.~\ref{fig:kinematics_all} summarizes the time traces of velocity and acceleration obtained for all impact Reynolds numbers for the two spheres $D=6$ and 10~mm:  Fig.~\ref{fig:kinematics_all}(a) and (b) are for $\rho^*=2.16$, (c) and (d) for $\rho^*=3.26$. This data clearly shows that the sphere deceleration decreases to very low values following cavity pinch-off, suggesting that the two final terms in (\ref{eq:Force_evaluation}), to first order, can be neglected after pinch-off. 
Thus, after pinch-off, the last two terms and the surface tension force in (\ref{eq:Force_evaluation}) are eliminated and the force balance can be written as 
\begin{equation}
\label{eq:force_balance}
    ma_s = F_{\text{G}}- F_{\text{H}} - F_{\text{B}},
\end{equation}

\noindent from which the hydrodynamic force $F_\mathrm{H}$ can be computed if all other terms are measured or can be computed. 

Lighter spheres with an attached air cavity reach a zero descent velocity and begin ascending, due to the dominance of buoyancy over gravity. Thus, this ascending trajectory behavior is expected under the condition
\begin{equation}
    \label{eq:ascending condition}
    \frac{F_\mathrm{B}}{F_\mathrm{G}}= \frac{\rho_lgV_\mathrm{Disp}}{m_sg}=\frac{V^*_\mathrm{Disp} }{\rho^*} >  1  \quad\quad \mathrm{or}\quad\quad V^*_\mathrm{Disp}>\rho^* ,
\end{equation}
a condition which can be monitored throughout the  trajectory following all  pinch-off events, to predict whether a sphere will continually descend or eventually ascend, i.e. terminally descending or ascending.
The dimensionless terminal velocity is obtained by equating the net body forces with the hydrodynamic force, leading to
\begin{equation}
    v_t^* = \frac{1}{v_i}\sqrt{ \frac{4 (\rho^* - 1)gD }{3C_D} }.
    \label{eq:terminal velocity}
\end{equation}
Since the sphere is neither accelerating nor decelerating when descending at the terminal velocity (steady flow), the hydrodynamic force has been expressed using a drag coefficient with the value  $C_D=0.5$, since the minimum Reynolds number experienced by any of the terminally descending spheres was approximately 1,200, i.e. in the Newtonian range.

\begin{figure}[t]
%\centering
\includegraphics[width=0.85\textwidth]{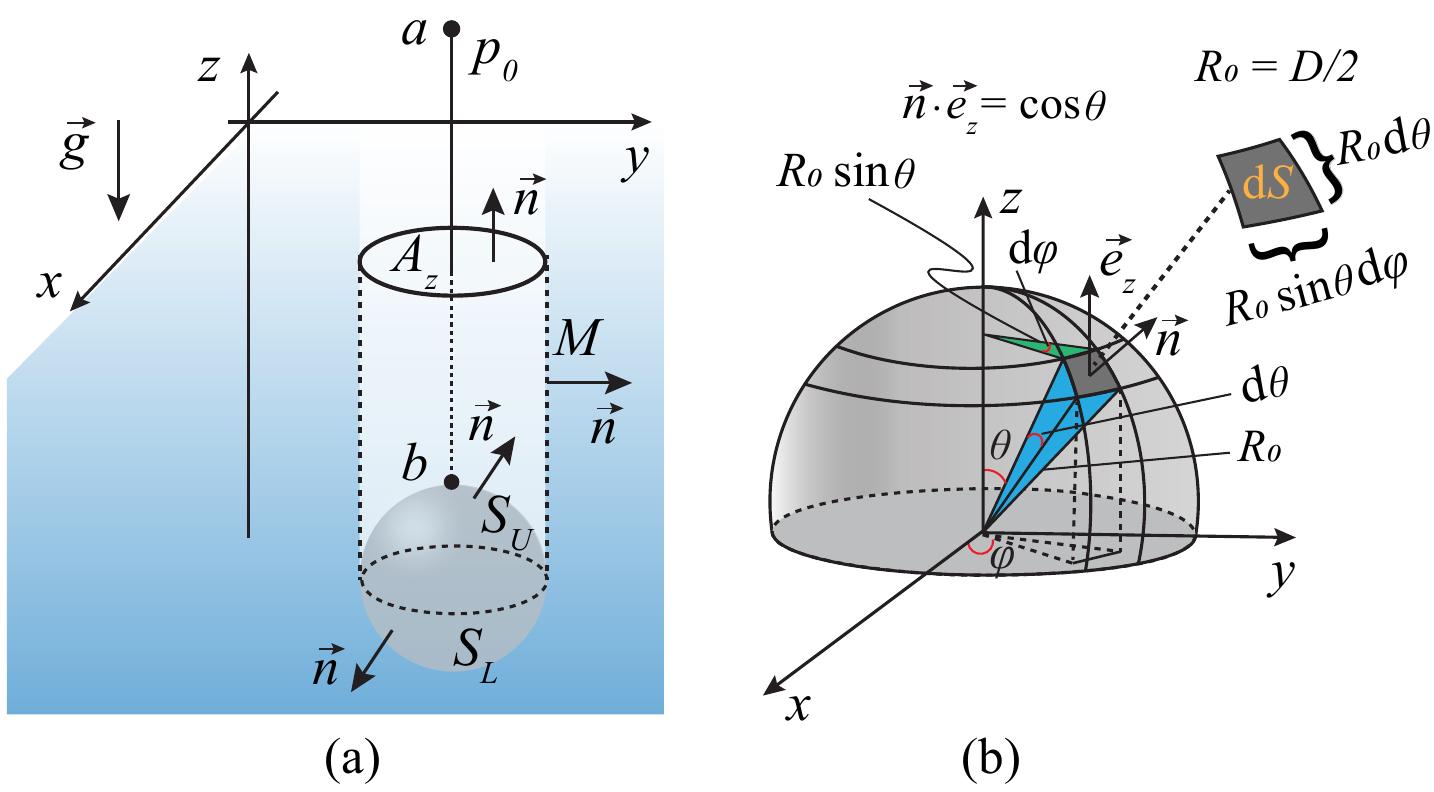}
\caption{(color online) Computation of hydrostatic force acting on the sphere before pinch-off. (a) Nomenclature for computation of the force acting on the lower hemisphere ($F_\mathrm{SL}$). (b) Nomenclature for computation of the force acting on the upper hemisphere ($F_\mathrm{SU}$). } 
\label{fig:buoyancy_with_cavity}
\end{figure}

However, before pinch-off, the body is not completely submerged and Archimedes principle is not strictly applicable; hence, a more detailed computation of the hydrostatic forces acting on the sphere is necessary. The approach is outlined in Fig.~\ref{fig:buoyancy_with_cavity}, which represents the situation after the sphere has submerged at least one diameter ($z_\mathrm{i}>D$), i.e. following the initial submergence and wetting of the sphere, but before pinch-off. 
For the present purposes, the assumption is made that the three-phase contact line is fixed at the equator and the forces acting on the sphere are the hydrostatic pressure force acting upward on the wetted lower surface ($F_\mathrm{SL}$), the air pressure acting downward over the upper surface ($F_\mathrm{SU}$) and the surface tension force acting upward at the equator, $F_\gamma=\pi D \gamma$, where $\gamma$ is the surface tension ($\gamma=0.072$~N/m). The viscous shear forces are neglected, since these are accounted for in the hydrodynamic force, $F_\mathrm{H}$. The net vertical hydrostatic force acting on the sphere before pinch-off then takes the form
\begin{equation}
    \label{eq:net_force}
    F_\mathrm{N}=F_\mathrm{SL}-F_\mathrm{SU}+F_\gamma.
\end{equation}
The vertical force acting on the lower hemisphere is given by \cite{spurk2007fluid}
\begin{equation}
    \label{eq:force_lower_surface}
    F_{SL}=\rho_l g V_e +p_0 A_z ,
\end{equation}
where $p_0$ is the pressure at the free surface and  $V_e = \pi D^3/12 - (z_i+D/2)\pi D^2/4$ is the ersatz volume consisting of the sphere and the virtual cylinder $M$ extending above the sphere from its equator to the surface (see Fig.~\ref{fig:buoyancy_with_cavity}(a)). Note that the $z_i$ coordinate of the sphere extracted from the videos at time step $i$ corresponds to the bottom of the sphere; hence, the $(z_i+D/2)$ factor in the last term of $V_e$. The area $A_z=\pi D^2/4$ is the projection of the sphere upwards onto the free surface.

The vertical force on the upper hemisphere is given by
\begin{eqnarray}
    \label{eq:S_U_area_integral}
    \vec{F}_{SU}\cdot \vec{e}_\mathrm{z}=
    F_{SU}&=&  -\iint\limits_{(S_U)} p\,\vec{n}\cdot \vec{e}_\mathrm{z}dS
    =-\int_0^{2\pi}\int_0^{\pi/2}p\,\vec{n}\cdot \vec{e}_\mathrm{z}\,R_0 \,\mathrm{d}\theta\, R_0\sin{\theta}\, \mathrm{d}\varphi \nonumber \\
    &=&- R_0^2 p\,\int_0^{2\pi}\left[\int_0^{\pi/2}\cos{\theta}\sin{\theta\mathrm{d}\theta}\right]\mathrm{d}\varphi
    =-\frac{\pi D^2 p}{4} ,
\end{eqnarray}
with $\vec{e}_\mathrm{z}$ being the unit vector in the z direction, $R_0$ the radius of the sphere and  $p$ being the air pressure acting on the surface (see Fig.~\ref{fig:buoyancy_with_cavity}(b), note $\vec{n}\cdot \vec{e}_\mathrm{z}=\cos{\varphi}$ ). The pressure $p$ can be estimated using the Bernoulli equation (assuming no losses) applied between the points $a$ and $b$ along a streamline extending from far above the surface (zero velocity) to the apex of the sphere, traveling at a velocity of $v_\mathrm{s}$, yielding
\begin{equation}
    \label{eq:bernoulli}
    p=p_0-\frac{1}{2}\rho_\mathrm{a}v^2_\mathrm{s} ,
\end{equation}
where $\rho_a$ is the density of the air. 

A final expression for the net force acting on the sphere before pinch-off and for $z_\mathrm{i}>D$ becomes
\begin{eqnarray}
    \label{eq:F_N}
    F_\mathrm{N}&=&\rho_l g \left(\frac{\pi D^3}{12} - \frac{\pi D^2}{4}(z_i+D/2)\right) +p_0 \frac{\pi D^2}{4} -\frac{\pi D^2}{4} \left(p_0-\frac{1}{2}\rho_\mathrm{a}v^2_\mathrm{s}\right)+\pi D\gamma   \nonumber \\
    &=& \rho_l g \left(\frac{\pi D^3}{12} - \frac{\pi D^2}{4}(z_i+D/2)\right)+ \frac{\pi D^2}{8}\rho_\mathrm{a}v^2_\mathrm{s} +\pi D\gamma .
\end{eqnarray}

 Given that $\rho_\mathrm{a}<<\rho_\mathrm{l}$, the penultimate term in the above equation is at least one order of magnitude smaller than the last term and can be neglected.  The final term in the above equation is of the order $10^{-3}$~N, which is approx. one order of magnitude less than the first term, but comparable to $F_\mathrm{G}$ for the small light spheres; hence, this term cannot always be neglected. Note, that  before pinch-off, $F_\mathrm{N}$ increases proportional to the depth of the sphere, $z_i$.
  
The hydrodynamic force acting on the sphere ($F_\mathrm{H}$)  can then be computed as:
\begin{equation}
    \label{eq:hydrodynamic_force}
    F_\mathrm{H}= \left\{
    \begin{array}{r@{\quad:\quad}l}
       F_\mathrm{G}-F_\mathrm{N}-ma_\mathrm{s} & \mbox{before pinch-off}\\
      F_\mathrm{G}-F_\mathrm{B}-ma_\mathrm{s} & \mbox{after pinch-off}
    \end{array}
    \right.
\end{equation}
whereby before pinch-off, $F_\mathrm{H}$ will also include any contributions from the added mass or history term in Eq.~(\ref{eq:Force_evaluation}), whereas after pinch-off, these terms will become negligible due to the low value of deceleration (see Fig.~\ref{fig:kinematics_all}).

\section{\label{sec:results}Results}
\subsection{Primary pinch-off}
\label{sec:primary}

The collapse of an empty cavity within an infinite liquid medium was first analyzed by Rayleigh~\cite{rayleigh1917viii} in the context of cavitation and bubble dynamics. The generalized Rayleigh–Plesset equation \cite{plesset1977bubble} describes the dynamics of a spherical bubble in a liquid under varying pressure conditions and is generally used to model cavity collapse. It is given by: 
\begin{eqnarray}
      \underbrace{ \rho_l \left( R \ddot R +\frac{3}{2} \dot R^2\right)}_\text{Inertial response}=  
      \underbrace{P_B (t) - P_\infty (t)}_\text{Pressure difference}
      - \underbrace{4\mu \frac{\dot R}{R}}_\text{Viscous}
      - \underbrace{2\frac{\gamma}{R}}_\text{Capillary}.
      \label{eq:General_Rayleigh} 
\end{eqnarray}

The left-hand side of Eq.~\ref{eq:General_Rayleigh} represents the inertial response of the bubble wall, while the right-hand side includes the driving pressure difference, viscous damping, and surface tension forces, respectively. Here, \( R \) denotes the bubble radius, and \( \dot{R} \) and \( \ddot{R} \) represent the first and second time derivatives of the radius, corresponding to the velocity and acceleration of the bubble wall, respectively. \( P_B \) is the pressure inside the bubble, and \( P_\infty \) is the ambient pressure far from the bubble in the liquid.

The entry cavity formation and its radial evolution, studied by \citet{Duclaux2007DynamicsCavities}, involves solving the Rayleigh–Plesset equation while considering the effects of gravity through the generalized Bernoulli equation. In their formulation, the minimum radius of the cavity, as expressed in their Eq.~(4.24), reaches zero at the moment of pinch-off. It has been shown in their analysis that the dimensionless cavity depth scales with the impact Froude number \(\mathrm{Fr}_i\). 

The generalized Rayleigh-Plesset equation assumes a purely radial expansion of the cavity. Neglecting viscous and surface tension forces, the governing equation for the cavity radius, \( R \), based on the hydrostatic pressure as a function of vertical depth \( z \), is given by \citet{Duclaux2007DynamicsCavities}:
\begin{equation}
    R \ddot{R} + \frac{3}{2} \dot{R}^2 = -gz.
    \label{eq:Cavity_evolution}
\end{equation}
The equation (12) is approximated as 
\begin{equation}
    \ddot{R^2} \approx -gz.
    \label{eq:approx_Cavity_evolution}
\end{equation}
An approximate analytical solution for the cavity shape evolution over time can be obtained by integrating Eq.~(\ref{eq:approx_Cavity_evolution}) twice using the initial conditions \( R(t = 0) = R_0 \) and \( \dot{R}(t = 0) = \sqrt{\alpha} U \). Here, $R_0$ is the radius of the sphere, \( \alpha \) is a constant with a value smaller than 1, and the square of the cavity radial velocity at the minimum radius is related to the impact velocity by \( \dot{R}^2 = \alpha U^2 \). Ultimately, the dimensionless minimum radius, \( \bar{R}_{\text{min}} \), where \( \bar{R}_{\text{min}} = R_{\text{min}} / R_0 \), is obtained as follows:

\begin{equation}
    \bar{R}^2_{\text{min}} = 1 + \frac{2\sqrt{2 \alpha \frac{z}{D}}}{9} \left[ 3 + 2 \sqrt{1-\frac{3 \mathrm{Fr}_i^2 \sqrt{\alpha}}{2(\frac{z}{D})^2}} \right] - \frac{16 \mathrm{Fr}_i^2}{27} \left( \frac{z}{D} \right)^3 \left[ 1 + \sqrt{1-\frac{3 \mathrm{Fr}_i^2 \sqrt{\alpha}}{ 2(\frac{z}{D})^2}} \right].
    \label{eq:minimum_radius}
\end{equation}

For values of \(\frac{z}{D} > 1/2\), where \(\bar{R}_{min} = 0\), the relationship between the dimensionless depth and the Froude number is determined from the (\ref{eq:minimum_radius}) as:
\begin{equation}
    \frac{z}{D} = \left( 2 \mathrm{Fr}_i^2\sqrt{\alpha} \right)^{\frac{1}{2}}.
    \label{eq:eta}
\end{equation}

In the regime where \(\mathrm{Fr}_i < 26\), the depth of the cavity increases linearly with the impact Froude number, implying a linear relationship between the depth and the impact velocity. The pinch-off location occurs at at approximately 50\% of the cavity depth. Given that the depth is approximately linearly proportional to the impact velocity (\(z = U t\)), the pinch-off time also varies linearly with the pinch-off depth. Consequently, the cavity length, pinch-off length, and pinch-off time all show a linear dependence on  $\mathrm{Fr}_i$.

\begin{figure}[htbp!]
    \begin{subfigure}[t]{0.42\textwidth}
      \includegraphics[width=\textwidth]{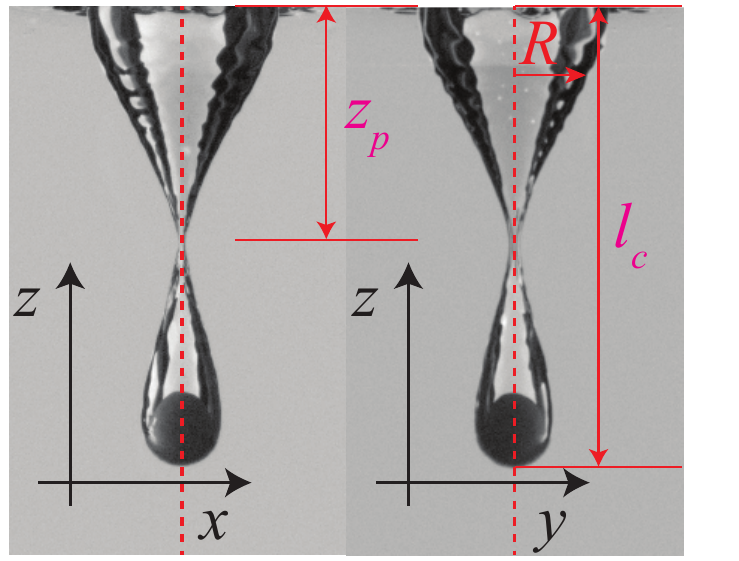}
      \caption{}
    \end{subfigure}
    \hspace{0.03\textwidth}
    \begin{subfigure}[t]{0.42\textwidth}
      \includegraphics[width=\textwidth]{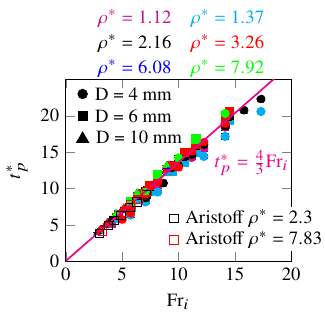}
      \caption{}
    \end{subfigure}
    \vspace{0.5em}
    \begin{subfigure}[t]{0.4\textwidth}
      \includegraphics[width=\textwidth]{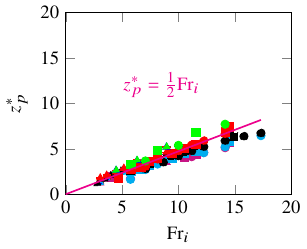}
      \caption{}
    \end{subfigure}
    \hspace{0.02\textwidth}
    \begin{subfigure}[t]{0.4\textwidth}
      \includegraphics[width=\textwidth]{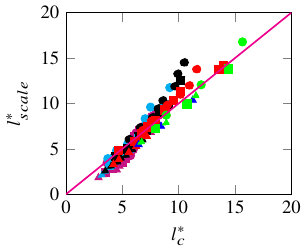}
      \caption{}
    \end{subfigure}
    \caption{(color online) Pinch-off characteristics of a superhydrophobic sphere entering a deep pool: (a) Dimensions of an axisymmetric cavity observed using two high-speed cameras at the pinch-off. The influence of $\mathrm{Fr}_i$ on the dimensionless pinch-off time $(t_p^*)$, and dimensionless pinch-off depth $(z^*_p)$ across various impacting conditions for four of the investigated densities is provided in (b) and (c), respectively. (d) Comparison of experimental dimensionless cavity length ($l_c^*$) with the scaled ($l_{scale}^*$) values obtained using Eq.~\ref{eq:l_c_nd}. The measurement uncertainties in (b)–(d) are on the order of the symbol size. The open square symbols in (b) correspond to the $t_p^*$ reported by~\citep{Aristoff2010TheSpheres}, plotted against the $\mathrm{Fr}_i$. The legend in (b) applies to (c) and (d).}
    \label{fig:pinch_off_regime}
\end{figure}

Although the present work does not focus on modeling the hydrodynamics of primary pinch-off, comparisons will be drawn to existing data, starting with Fig.~\ref{fig:pinch_off_regime} in which the dimensionless pinch-off time $(t_p^*)$, dimensional pinch-off depth $(z_p^*)$ and total cavity length $(l_c^*)$ are shown as a function of $\mathrm{Fr}_i$ in Fig.~\ref{fig:pinch_off_regime}(b)-(d), respectively for various densities. The results for $\rho^* = 2.16$ and $\rho^* = 7.92$ exhibit similarities with, and extend those observed in \citet{Aristoff2010TheSpheres}, particularly in terms of the linear relationship with $\mathrm{Fr}_i$. Upon impacting the air-water interface, the higher density sphere carries more momentum and penetrates further in the time required for the pinch-off. This leads to larger values of the total cavity length $l_c^*$. This is also reflected in the pinch-off position $z_p^*$, whereas the dimensional pinch-off time ($t_p$) is presented in the Appendix (see Fig.~\ref{fig:Dimensional_pinch_off_time}(a)), is more or less independent of the sphere density. Note that the dimensionless pinch-off depth $(z_p^*)$ is consistently 40-60\% of $(l_c^*)$.

From Fig.~\ref{fig:pinch_off_regime}(b) it is apparent that the pinch-off occurs at a constant dimensionless time, regardless of the impact conditions or sphere size and density. The experimental data lead to the linear relation $t_p^*=4\textrm{Fr}_\mathrm{i}/3 $ yielding for the absolute pinch-off time
\begin{equation}
    \label{eq:t_Fr}
    t_\mathrm{p}=\frac{4}{3}\sqrt{\frac{D}{g}}.
\end{equation}
This relation is confirmed by the measured $t_p$ values shown in the Appendix, Fig.~\ref{fig:Dimensional_pinch_off_time}(a), and indicates that the absolute pinch-off time does not depend on the impact velocity or density of the sphere, but only on the sphere diameter.  This is intuitively plausible, since the pinch-off is a phenomenon dominated primarily by capillary forces; the impact velocity only serves to determine the size of the cavity before pinch-off.

To estimate the size of the cavity at pinch-off, the distance the sphere travels in $t_p$, denoted $l_c$ in Fig.~\ref{fig:pinch_off_regime}(d) is first computed. To the first order, the acceleration of the sphere before pinch-off can be approximated as
\begin{equation}
    \label{eq:acceleration}
    a_s=g-F_\mathrm{N}/\rho_s V_s=g-\frac{5g}{4\rho^*}-\frac{3g}{2\rho^*}\left(\frac{z_i}{D}\right),
\end{equation}
whereby only the leading terms terms of $F_\mathrm{N}$ have been included and forces due to added mass and the history term in Eq.~(\ref{eq:F_N}) have been neglected. The distance traveled by the sphere is then given by
\begin{eqnarray}
    \label{eq:distance}
    z(t) &= &z_0 + v_0 t + a_\mathrm{s}t^2/2 \nonumber \\ 
         &=& v_\mathrm{i}t+\left[g-\frac{5g}{4\rho^*}-\frac{3g}{2\rho^*}\left(\frac{z_\mathrm{i}}{D}\right)\right]\frac{t^2}{2}.
\end{eqnarray}
Inserting $t_\mathrm{p}$ and solving for $z$ yields a first-order estimate for $l_\mathrm{c}$
\begin{equation}
    \label{eq:l_c}
    l_\mathrm{scale}=\frac{\frac{4v_\mathrm{i}}{3}\sqrt{\frac{D}{g}}+\left[ 1-\frac{5}{4\rho^*} \right]\frac{16D}{18}}{\left[1+\frac{4}{3\rho^*}\right]},
\end{equation}

\begin{equation}
    \label{eq:l_c_nd}
    l_\mathrm{scale}^*=\frac{\frac{4v_\mathrm{i}}{3}\sqrt{\frac{1}{Dg}}+\left[ 1-\frac{5}{4\rho^*} \right]\frac{16}{18}}{\left[1+\frac{4}{3\rho^*}\right]}.
\end{equation}

Similarly, $z^*_p$ can be approximated from Fig.~\ref{fig:pinch_off_regime}(c) with the expression $z^*_p=0.5$~Fr$_i$, leading to
\begin{equation}
    \label{eq:z_p}
    z_p=\frac{v_i}{2g}.
\end{equation}

A first-order estimate of the cavity height immediately following pinch-off (see Fig.~\ref{fig:pinch_off_regime}) is given by
\[
H = l_c - z_p - \frac{D}{2}.
\]

\begin{figure}
    \centering
    \includegraphics[width=0.8\textwidth]{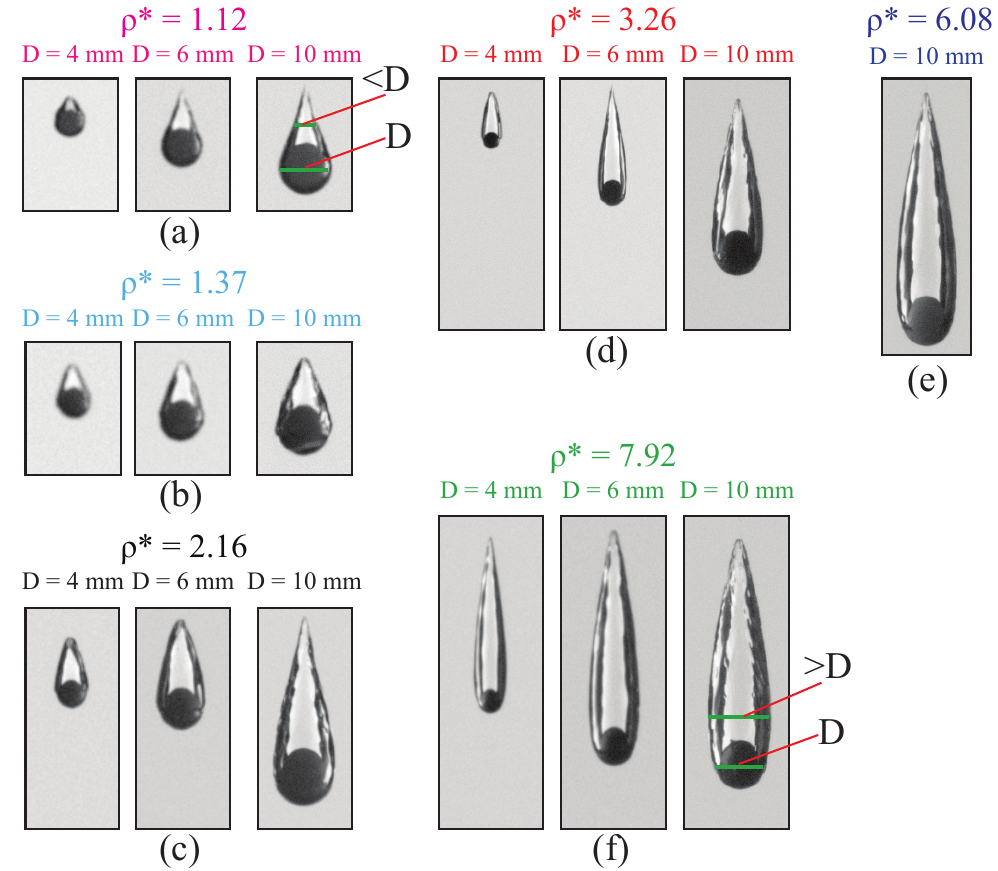}
    \caption{(color online) Cavity shapes captured after primary pinch-off for spheres with (a) $\rho^* = 1.12$, (b) $\rho^* = 1.37$, (c) $\rho^* = 2.16$, (d) $\rho^* = 3.26$, (e) $\rho^* = 6.08$, and (f) $\rho^* = 7.92$. Each subfigure shows the cavity formed by 4~mm (left, $\mathrm{Fr}_i$ = 14), 6~mm (middle, $\mathrm{Fr}_i$ = 11.5), and 10~mm (right, $\mathrm{Fr}_i$ = 9) spheres. For $\rho^* = 6.08$, only the 10~mm case ($\mathrm{Fr}_i$ = 9) is shown.}
    \label{fig:After_pinchoff}
\end{figure}

For lighter spheres (\( \rho^* = 1.12 \) and \( 1.37 \)), the volume displaced by the cavity after pinch-off can be approximated as a conical volume. In this case, the total displaced volume is considered as the sum of the conical gas cavity and a hemispherical cap at the base as illustrated in Fig.~\ref{fig:After_pinchoff}(a) and (b).

For heavier spheres (\( \rho^* \geq 2.16 \)) (shown in Figs.~\ref{fig:After_pinchoff}(c-f)), the maximum cavity diameter exceeds the diameter of the impacting sphere (\cite{Aristoff2009WaterSpheres,Vakarelski2017Self-determinedCavities}). Although the cavity shape immediately after the pinch does not strictly resemble a perfect cylinder (see Fig.~\ref{fig:After_pinchoff}), a modified height term, denoted here as the \textit{effective height}, is introduced in the scaling expression. This adjustment allows the simplified geometry (combined cylindrical gas cavity volume and hemispherical cap at the base) to better represent the displaced volume.

To represent the cavity shape consistently across all density ratios, a model for the scaled displaced volume is proposed as follows:
\begin{equation}
V_{\text{Disp}}^{\text{scaled}\,^*} =
    \begin{cases}
    \displaystyle \frac{\pi D^2}{12} \left(H+\frac{D}{2}\right), & \rho^* \leq 1.37 \quad \text{(conical)} \\[10pt]
    \displaystyle \frac{\pi D^2}{12} \left(2H - \frac{D}{2}\right), & \rho^* \geq 2.16 \quad \text{(cylindrical)} \\[10pt]
\end{cases}
\label{eq:cavity_approximation}
\end{equation}

\begin{figure}[t]
%\centering
\includegraphics[width=0.5\textwidth]{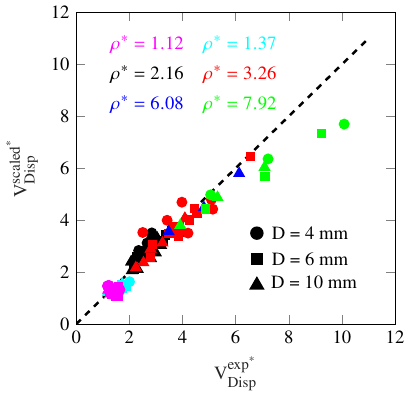}
\caption{(color online) Comparison between scaled model-predicted dimensionless displaced volumes ($V_{\text{Disp}}^{\text{scaled}\,^*}$) predicted by the model and experimentally measured dimensionless displaced volumes ($V_{\text{Disp}}^{\text{exp}\,^*}$) following the primary pinch-off. } 
\label{fig:V_Disp_Scale_Exp}
\end{figure}

 By introducing a correction through the \textit{effective height} for the heavier spheres \( \rho^* \geq 2.16 \), which accounts for deviations in the actual cavity profile,  the simplified geometric approximation (cylinder with hemisphere at the base) captures the experimentally measured cavity volume with good accuracy, as shown in Fig.~\ref{fig:V_Disp_Scale_Exp}. Larger deviations from this Froude number scaling occur only for the heavier and smaller spheres. This is presumably due to the increased absolute uncertainty in determining the gas cavity volume from the videos, which is then greater amplified in dimensionless form with smaller spheres.

\begin{figure}
    \centering
    \begin{subfigure}{0.41\textwidth}
        \centering
        \includegraphics[width=\linewidth]{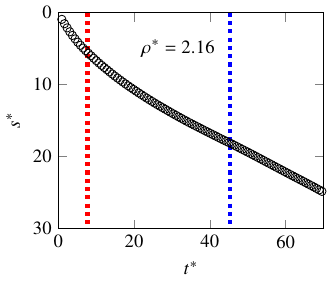}
        \caption{}
    \end{subfigure}
    \hspace{0.05\textwidth}
    \begin{subfigure}{0.42\textwidth}
        \centering
        \includegraphics[width=\linewidth,trim=0 0 0 1.5,clip]{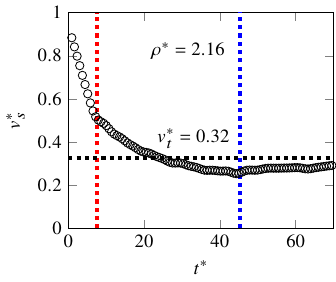}
        \caption{}
    \end{subfigure}
    
    \begin{subfigure}{0.42\textwidth}
        \centering
        \includegraphics[width=\linewidth,trim=0 3 0 0,clip]{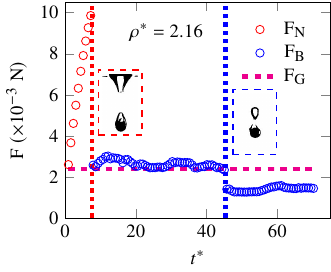}
        \caption{}
    \end{subfigure}
    \hspace{0.05\textwidth}
    \begin{subfigure}{0.42\textwidth}
        \centering
        \includegraphics[width=\linewidth,trim=0 0 0 2,clip]{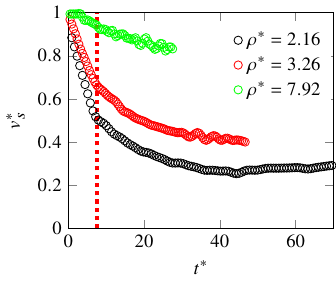}
        \caption{}
    \end{subfigure}
    \caption{(color online) The influence and magnitude of hydrostatic forces on the sphere ($\rho^* = 2.16$) trajectory as a function of dimensionless time. (a) Dimensionless sphere path $s^*$, (b) dimensionless velocity $v_s^*$, (c) forces acting before ($F_\mathrm{N}$) and after ($F_\mathrm{B}$) the primary pinch-off, and (d) dimensionless sphere velocity for various dimensionless densities. Vertical dashed red and blue lines designate the dimensionless time of primary and secondary pinch-off, respectively. Data is for $D=6$~mm and Re$_i=9,400$. The horizontal dashed black line in (b) marks the terminal velocity according to Eq.~(\ref{eq:terminal velocity}). The horizontal dashed magenta line in (c) marks $F_\mathrm{G}$ allowing the condition in Eq.~(\ref{eq:ascending condition}) to be directly evaluated. The inset in (c) shows captured images immediately following the primary and secondary pinch-off events, highlighted by red and blue dashed boxes, respectively. }
    \label{fig:Buoyancy}
\end{figure}

The influence of the primary pinch-off on the trajectory and velocity is illustrated in Fig.~\ref{fig:Buoyancy}.  In this figure,  graph (b) shows the dimensionless velocity of the sphere over dimensionless time and noteworthy is a distinct kink and decreased deceleration rate at the time of pinch-off. This is clearly associated with the hydrostatic force jumping from $F_\mathrm{N}$ to the much lower $F_\mathrm{B}$ at the moment of pinch-off, as shown in graph (c). In graph (d) similar velocity curves for other dimensionless sphere densities are shown. It is evident from this last graph that the influence of hydrostatic forces is much stronger for the lighter spheres due to their lower inertial force; for heavier spheres there is no distinguishable kink.

In graph (c) a horizontal dashed line at $F_\mathrm{G}$ has been added, allowing condition (\ref{eq:ascending condition})  to be directly evaluated. For this particular case ($D=6$~mm, Re$_i=9,400$) $F_\mathrm{B}$ lies slightly above $F_\mathrm{G}$ immediately after the primary pinch-off, indicating that the sphere and attached cavity would eventually begin to rise. However, after a second pinch-off at approx. $t^*=45$, $F_\mathrm{B}<F_\mathrm{G}$, indicating that the sphere and remaining cavity will continue descending. This change in buoyancy after the secondary pinch-off is also seen in graph (b), where the slope of $v^*_\mathrm{s}$ changes from negative (decelerating) to positive (accelerating) at $t^*=45$, subsequently approaching the terminal velocity. It should be noted that the terminal velocity according to Eq. \ref{eq:terminal velocity} may no longer be strictly valid, as the assumed $C_\mathrm{D}=0.5$ is for a sphere, not a sphere with an attached air cavity. The heavier spheres do not approach their terminal velocity within the observational field of view.

Although the total cavity length at pinch-off is different for different density spheres, the dimensional primary pinch-off time remains virtually unchanged, as shown in the Appendix (see Fig.~\ref{fig:Dimensional_pinch_off_time}(a)). The total cavity length reaches higher values for higher dimensionless densities before pinch-off; following pinch-off the buoyancy to gravity force ratio ($V_\mathrm{Disp}^*$, Eq.~\ref{eq:ascending condition}) decreases with increasing sphere density, exhibiting  values less than unity for the heaviest sphere. 

\begin{figure}
    \begin{subfigure}{0.55\linewidth}
        \centering
        \includegraphics[width=\linewidth]{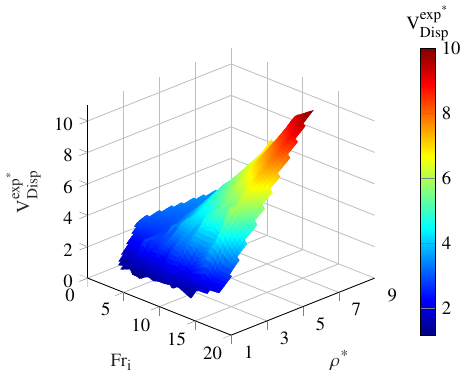} 
    \end{subfigure}
    \hspace{0.02\linewidth} 
    \begin{subfigure}{0.36\linewidth}
        \includegraphics[width=\linewidth]{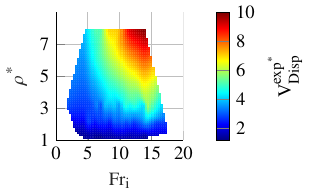} 
        \vspace{0.01\linewidth} 
        \hspace{-9.5mm}
        \includegraphics[width=\linewidth]{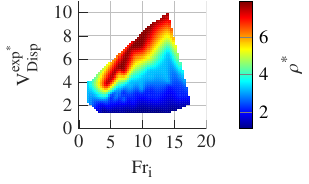} 
    \end{subfigure}
    \caption{(color online) Three-dimensional surface plot of the dimensionless displaced volume ($V_{\text{Disp}}^{\text{exp}^*}$) as a function of the (\( \mathrm{Fr}_i \)) and (\( \rho^* \)). The plot illustrates that both increasing impact inertia and density ratio lead to greater displaced volume. To enhance interpretability, two-dimensional projections of the surface are provided as insets. The top view inset (\( \rho^* \) vs. \( \mathrm{Fr}_i \)) illustrates the density dependent variation of $V_{\text{Disp}}^{\text{exp}^*}$, while the side view inset ($V_{\text{Disp}}^{\text{exp}^*}$) vs. \( \mathrm{Fr}_i \)) further emphasizes the monotonic rise in displaced volume with increasing \( \mathrm{Fr}_i \), especially at higher density ratios.}
    \label{fig:Contour3D_volumes}
\end{figure}

To further elucidate the combined effects of inertia and buoyancy on the $V_{\text{Disp}}^{\text{exp}^*}$, Fig.~\ref{fig:Contour3D_volumes} presents a three-dimensional surface plot. The surface demonstrates a smooth and monotonic increase in the dimensionless displaced volume ($V_{\text{Disp}}^{\text{exp}^*}$) with both (\( \mathrm{Fr}_i \)) and (\( \rho^* \)). At low values of these parameters, the $V_{\text{Disp}}^{\text{exp}^*}$ remains small, indicating minimal cavity formation. However, as either \( \mathrm{Fr}_i \) or \( \rho^* \) increases, there is a marked growth in $V_{\text{Disp}}^{\text{exp}^*}$, indicating that spheres with greater inertia or higher density relative to the fluid displace significantly more liquid.

To aid understanding of this coupled influence, two-dimensional projections are included as insets. The top view projection onto the \( \rho^* \)–\( \mathrm{Fr}_i \) plane illustrates the spatial distribution and density dependent variation of $V_{\text{Disp}}^{\text{exp}^*}$. The side view projection onto the $V_{\text{Disp}}^{\text{exp}^*}$–\( \mathrm{Fr}_i \) plane further highlights the consistent increasing trend in displaced volume with increasing \( \mathrm{Fr}_i \), across all density ratios.

\begin{figure}
    \centering
    \includegraphics[width=\linewidth]{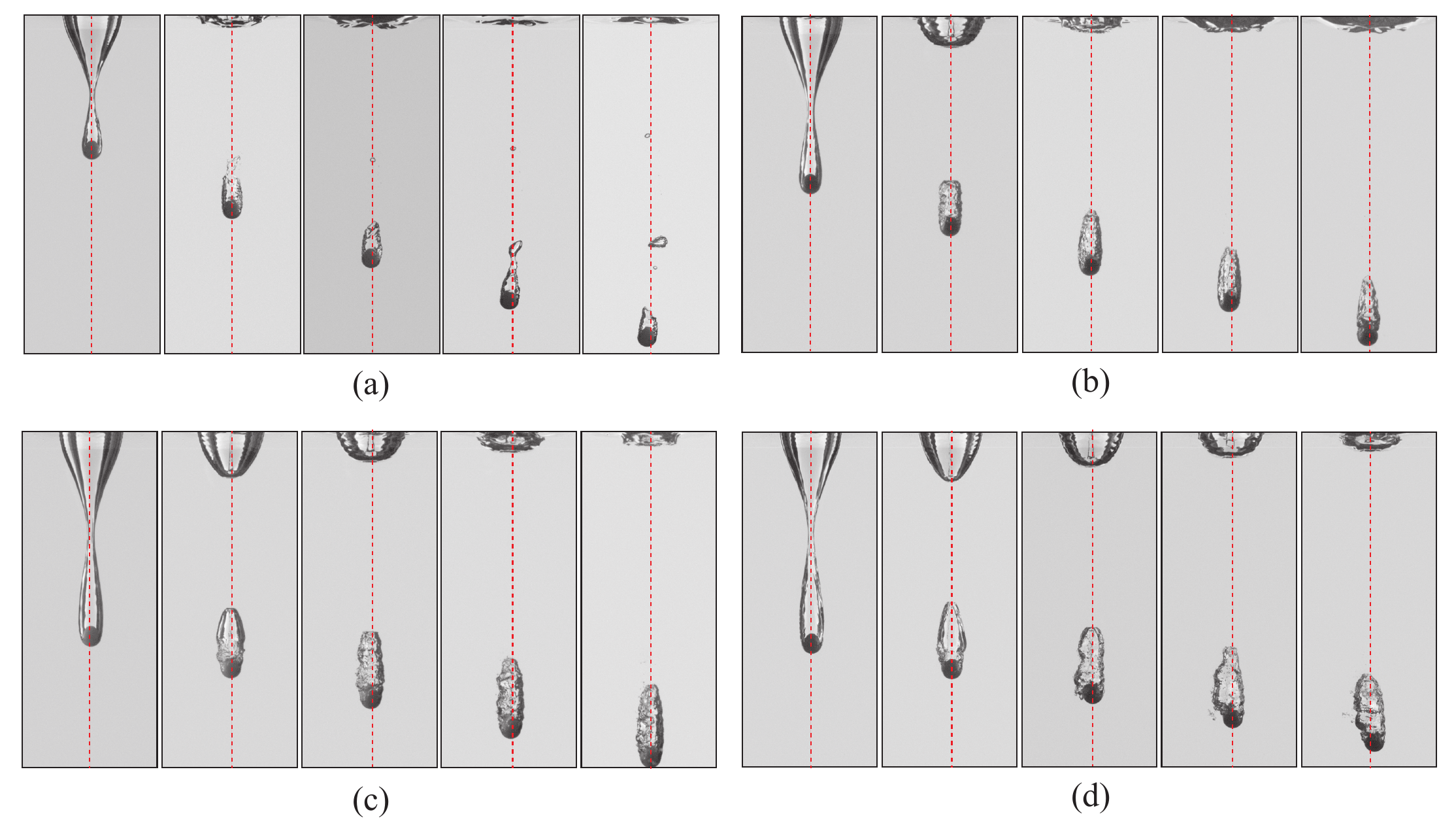}
    \caption{(color online) Series of images of a $D=10$~mm sphere impacting with $\mathrm{Re}_i = 31,500$ for various density ratios: (a) $\rho^*=2.16$, (b) $\rho^*=3.26$, (c) $\rho^*=6.08$, and (d) $\rho^*=7.92$. Each sequence begins with the moment of primary pinch-off, followed by the evolution of the air cavity. After pinch-off, heavier spheres ($\rho^* = 6.08$ and 7.92) shown in (c) and (d) exhibit strong fluctuations and asymmetry, whereas lighter spheres ($\rho^* = 2.16$ and 3.26) illustrated in (a) and (b) experience significantly lower fluctuations. The time interval between consecutive frames is 30 ms for (a), 15 ms for (b), 8 ms for (c), and 6 ms for (d). A red dashed vertical line in all images indicates the sphere impact location.}
    \label{fig:cavity fluctuations}
\end{figure}

\subsection{Secondary pinch-off}

\begin{figure}
    \centering
    \includegraphics[width=\linewidth]{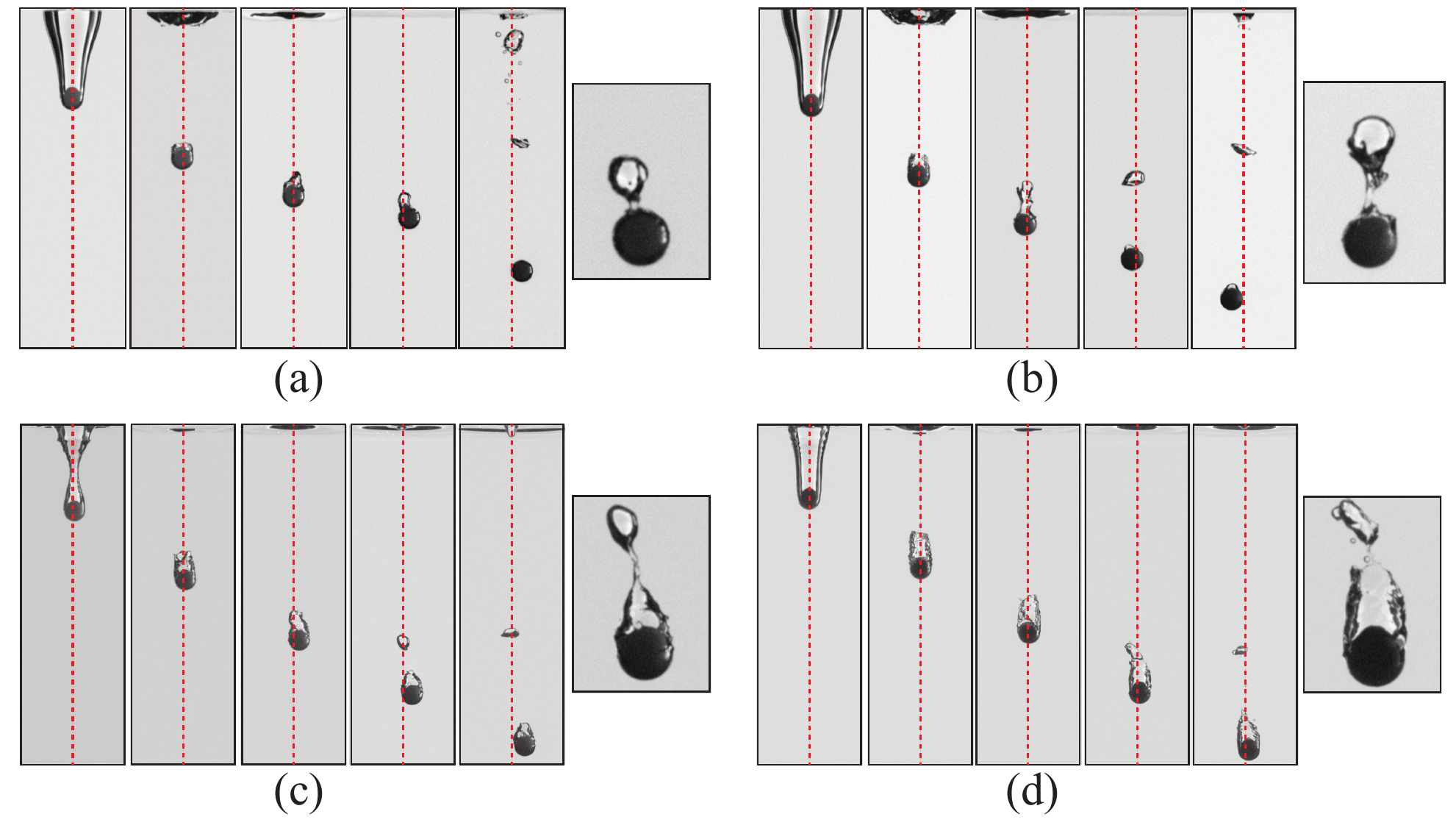}
    \caption{(color online) Image sequence of a $D=10$ mm sphere impacting the liquid surface at various density ratios: (a) $\rho^* = 1.12$, (b) $\rho^* = 1.37$ at $\mathrm{Re}_i = 31{,}500$, (c) $\rho^* = 2.16$, and (d) $\rho^* = 3.26$ at $\mathrm{Re}_i = 15{,}700$. Each sequence illustrates the cavity initiation, its subsequent evolution, and the occurrence of the secondary pinch-off, followed by rupture and bubble detachment. Insets on the right highlight the secondary pinch-off event in greater detail for each density ratio.}
    \label{fig:raw_secondary_pinch}
\end{figure}

Following the primary pinch-off, a secondary pinch-off occurs under most conditions, examples of which are illustrated in the still video sequences shown in Fig.~\ref{fig:raw_secondary_pinch} for the penetration of a 10~mm diameter sphere  for four different density ratios up to $\rho^*=3.26$. For spheres with high density ratios (\( \rho^* = 6.08 \) and \( 7.92 \)) and larger diameters the secondary pinch-off is not observed within the field of view. The corresponding dimensional pinch-off times ($t_{2p}$), which depend on both the sphere density ($\rho^*$) and $\mathrm{Fr}_i$, are presented in Appendix (see Fig.~\ref{fig:Dimensional_pinch_off_time}(b)). The secondary pinch-off can take different forms - either a cloud of smaller droplets (see Fig.~\ref{fig:raw_secondary_pinch}(a)) or a single bubble separating from the air cavity (see Fig.~\ref{fig:raw_secondary_pinch}(b)-(d)). For lighter spheres of smaller diameter, multiple pinch-offs can be observed. In any case, the displaced volume $V_{\text{Disp}}^{\text{exp}^*}$ reduces and often the ratio of buoyancy to gravity forces falls from above to below unity, resulting in a terminally descending trajectory.

\begin{figure}
    \centering
    \begin{subfigure}{0.44\textwidth}
        \includegraphics[width=\textwidth]{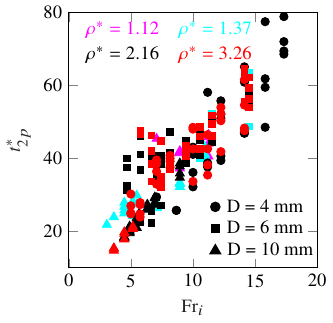}
        \caption{}
    \end{subfigure}
    \hspace{0.02\textwidth}
    \raisebox{-2mm}{%
        \begin{subfigure}{0.43\textwidth}
            \includegraphics[width=\textwidth]{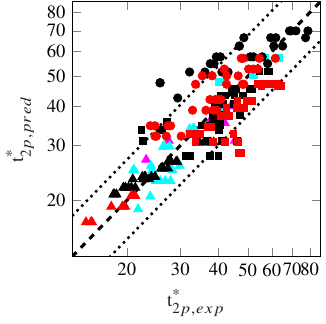}
            \caption{}
        \end{subfigure}%
    }
    \caption{(color online) (a) Experimentally measured dimensionless secondary pinch-off time, $t_{2p,exp}^*$, as a function of $\mathrm{Fr}_i$, for density ratios $\rho^* \leq 3.26$. (b) Predicted dimensionless secondary pinch-off time, $t_{2p,pred}^*$, from the regression $t_{2p,pred}^* = 0.345\,\mathrm{Fr}_i^{-0.45}\,\rho^{*-0.22}$ compared with experiments. The data clustering near the 1:1 line, within uncertainty bounds ($\pm 25\%$), demonstrates that the scaling captures the leading dependence of secondary pinch-off on impact inertia and density ratio on the convective timescale. The legends and colors in (a) also applicable to (b). }
    \label{fig:secondary_pinch_Fr}
\end{figure}

Figure~\ref{fig:secondary_pinch_Fr}(a) presents the experimentally measured dimensionless secondary pinch-off time, $t_{2p,exp}^*$, for density ratios $\rho^* \leq 3.26$, plotted against the $\mathrm{Fr}_i$. This time increases with $\mathrm{Fr}_i$, indicating that the impact inertia dominates over the gravity force, reducing the effectiveness of hydrostatic pressure in restoring the cavity. Consequently, the onset of secondary pinch-off is delayed, leading to systematically larger $t^{*}_{2p}$ at higher $\mathrm{Fr}_i$. A simple scaling argument supports this trend: the hydrostatic recovery time scales as $t \sim v_i/g$, yielding $t^{*} \sim v_i^2/(gD) = \mathrm{Fr}_i^2$, consistent with the observed increase of $t^{*}_{2p}$ with $\mathrm{Fr}_i$ (see Fig. \ref{fig:secondary_pinch_Fr}(a)).

Figure~\ref{fig:secondary_pinch_Fr}(b) indicates that the two-parameter regression 
\begin{equation}
t_{2p,pred}^* = 0.345\,\mathrm{Fr}_i^{-0.45}\,\rho^{*-0.22},
\label{eq:t2p}
\end{equation}
 predicts the measured $t_{2p}^*$ reasonably well.  At first glance, the negative exponent on $\mathrm{Fr}_i$ in the regression may appear inconsistent with the positive trend observed in Fig.~\ref{fig:secondary_pinch_Fr}(a). This apparent discrepancy is resolved by recognizing that the regression describes how the pinch-off time varies relative to the convective scale $D/v_i$. With increasing $v_i$ (and thus $\mathrm{Fr}_i$), the cavity becomes more elongated and the absolute delay before secondary pinch-off increases, producing the positive experimental trend in Fig.~\ref{fig:secondary_pinch_Fr}(a). However, the characteristic convective timescale $D/v_i$ decreases more rapidly with $v_i$, so that when expressed on this basis, the regression correctly identifies a weak negative dependence of $t_{2p,pred}^*$ on $\mathrm{Fr}_i$.  The weak negative dependence on $\rho^*$ further suggests that heavier spheres sustain elongated cavities and low pressure wake regions, which subtly shift the timing of secondary neck formation when viewed on the convective scale. Together, these observations underline that the secondary pinch-off arises from a balance between inertia driven cavity elongation, hydrostatic pressure recovery, and capillary stresses acting at the neck.

\subsection{\label{sec:force balance}Hydrodynamic force}
\begin{figure}[t]
\centering
\begin{subfigure}[b]{0.49\textwidth}
        \centering
        \includegraphics[width=\linewidth]{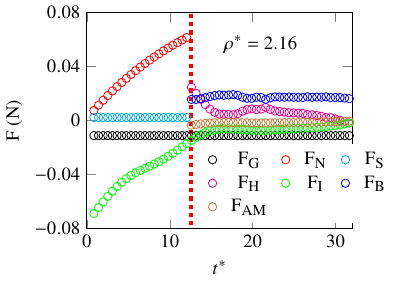}
        \caption{}
    \end{subfigure}
    %\hfill
    \begin{subfigure}[b]{0.49\textwidth}
        \centering
        \includegraphics[width=0.85\linewidth]{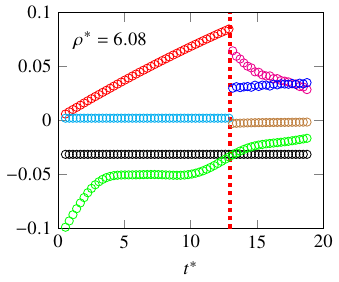}
        \caption{}
    \end{subfigure}
\caption{(color online) Instantaneous forces before and after the pinch-off of a 10\,mm diameter sphere impacting at $\mathrm{Re}_i = 31,500$ for (a) $\rho^* = 2.16$ and (b) $\rho^* = 6.08$. The red dashed vertical line indicates the dimensionless pinch-off time ($t_p^*$). The legend shown in (a) also applies to (b).} 
\label{fig:force_evolution}
\end{figure}

The time evolution of all force components is summarized in Fig.~\ref{fig:force_evolution} for 10~mm spheres of dimensionless density $\rho^*=2.16$ and 6.08 impacting at a Reynolds number of 31,500, whereby the hydrodynamic force $F_\mathrm{H}$ is the difference necessary to yield a sum of zero over all other forces. 
The net force before pinch-off, $F_\mathrm{N}$, is expected to include any viscous or pressure drag, inertia due to added mass, as well as forces arising from the history term in Eq.~(\ref{eq:Force_evaluation}). This force increases with sphere depth, as expressed in Eq.~(\ref{eq:F_N}). The changes in inertial force ($F_\mathrm{I}$) are initially significant immediately after impact, reflecting the rapid deceleration of the sphere. This force eventually vanishes once the sphere attains its terminal velocity, where the deceleration becomes negligible, as also shown in Fig.~\ref{fig:kinematics_all}. The gravitational force ($F_\mathrm{G}$) remains constant throughout the descent, completely independent of the sphere velocity. The added mass ($F_\mathrm{AM}$) contribution for a completely submerged sphere (with no air cavity) can be shown to be half the volume of the sphere times the density of the fluid, probably less for the more streamlined sphere with closed air cavity. However, before pinch-off, when the deceleration is still non-negligible and an open air cavity persists, there exist no reliable data or theory for the added mass. Thus, before pinch-off, it is not feasible to separate the individual force components contained in $F_\mathrm{N}$. After pinch-off, the added mass is computed using a value of $C_\mathrm{A}=0.5$ and the buoyancy force ($F_\mathrm{B}$) can be computed according to Archimedes' principle; thus, the hydrodynamic force $F_\mathrm{H}$ will ostensibly include only viscous shear and pressure forces, normally known as drag. 
The force due to surface tension $F_\mathrm{S}$ is only shown before pinch-off and, as expected, is very small. Similarly, the added mass force after pinch-off is very small, since the sphere is no longer undergoing significant deceleration. For the same reason, the history term has not been included in this diagram after pinch-off.

Two anomalies are now discussed. After pinch-off the buoyancy term shows a slight increase, more pronounced for the $\rho^*=6.08$ sphere. This is anomalous, since the displaced volume cannot change, there being no source of additional air after pinch-off. This behavior is attributed to the initial unsteadiness of the wake air cavity and the uncertain estimation of its volume from the videos. At longer times, the buoyancy value becomes constant, presumably because the cavity becomes more steady with time.

The second anomaly is the decrease in $F_\mathrm{H}$ after pinch-off, which represents primarily the drag. A small portion of this decrease can be attributed to the decrease in the added mass in the force balance, but this is minimal in comparison. The curve of $F_\mathrm{I}$ indicates some residual deceleration after pinch-off, but this is not expected to alter the drag noticeably. However, two further effects can be responsible for this decrease. For one, the shape of the air cavity can become more streamlined with time, leading to a lower drag coefficient. Secondly, during   pinch-off, a bi-directional gas flow arises at the neck, acting upwards above the pinch-off neck and below the neck, acting downward in the direction of the sphere. This gas flow will impart a dynamic pressure over the upper surface of the sphere ($S_\textrm{U}$), which exceeds the $p_0$ existing before necking begins. Furthermore, immediately after pinch-off, the gas cavity will exhibit an extremely small radius of curvature at its tail, leading to a high Laplace pressure in the gas cavity. The resulting pressure jump across the interface is with respect to the hydrostatic pressure at the pinch-off depth, which is already well above the atmospheric pressure acting on the upper sphere surface up to the pinch-off instant.  This higher static pressure acts to re-shape the wake air cavity, but will also impart a flow and corresponding pressure force on the sphere in the downward direction. Both this pressure force and the dynamic pressure due to the bi-directional gas flow result in a force acting on the upper surface of the sphere with the opposite sign as conventional drag. This would explain the change of effective drag expressed by $F_\mathrm{H}$ until the wake cavity shape relaxes.

\subsection{Terminally ascending or descending}

\begin{figure}
    \begin{subfigure}[t]{0.33\textwidth}
    \includegraphics[width=\linewidth]{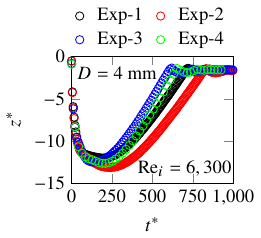}
    \caption{}
   \end{subfigure}
   \hfill
   \begin{subfigure}[t]{0.325\textwidth}
   \includegraphics[width=0.9\linewidth]{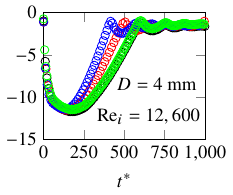}
 \caption{}
   \end{subfigure}
   \hfill
   \begin{subfigure}[t]{0.32\textwidth}
   \includegraphics[width=0.9\linewidth]{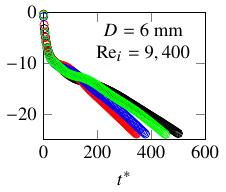}
 \caption{}
   \end{subfigure}
   \hfill
   \begin{subfigure}[t]{0.33\textwidth}
   \includegraphics[width=\linewidth]{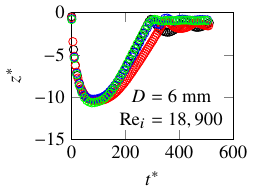}
 \caption{}
   \end{subfigure}
   \hfill
   \begin{subfigure}[t]{0.325\textwidth}
   \includegraphics[width=0.9\linewidth]{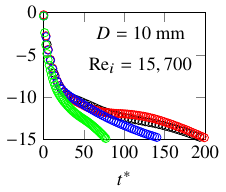}
 \caption{}
   \end{subfigure}
   \hfill
   \begin{subfigure}[t]{0.32\textwidth}
   \includegraphics[width=0.9\linewidth]{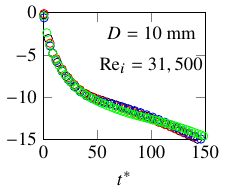}
 \caption{}
   \end{subfigure}
    \caption{(color online) Depth trajectories of a $\rho^* = 1.12$ sphere in repeated experiments. (a) $D =$ 4~mm, Re$_i$~=~6,300; (b) $D =$ 4~mm, Re$_i$~=~12,600; (c) $D =$ 6~mm, Re$_i$~=~9,400; (d) $D =$ 6~mm, Re$_i$~=~18,900; (e) $D =$ 10~mm, Re$_i$~=~15,700; (f) $D =$ 10~mm, Re$_i$~=~31,500. The legend in (a) is applicable to (b)-(f).}
    \label{fig:sphere 1.12}
\end{figure}

From the above discussion and according to condition (Eq.~\ref{eq:ascending condition}), a sphere will be terminally ascending or descending depending on whether $V_\mathrm{Disp}^*> \rho^*$ or $V_\mathrm{Disp}^*< \rho^*$ respectively, after all pinch-off events have occurred. Thus, the question of terminally ascending or descending is intimately coupled to the probability of pinch-off events, especially secondary pinch-offs, since immediately after the primary pinch-off the buoyancy almost always exceeds the gravitational force, but the sphere continues descending  because of its inertia. Furthermore, for a given sphere diameter and impact Reynolds number, the probability of it terminally ascending will be higher as the sphere mass decreases, and this is experimentally observed when examining the trajectory of the smallest and lightest spheres, e.g. $\rho^*=1.12$, shown in Fig.~\ref{fig:sphere 1.12}. The $D=4$~mm sphere always ends up ascending back to the free surface and the $D=10$~mm sphere is always terminally descending. For the \( D = 6\,\text{mm} \) sphere, a transition in terminal behavior is observed with varying impact Reynolds number. At \( \mathrm{Re}_i = 9,400 \) the sphere exhibits terminal descending motion (see Fig.~\ref{fig:sphere 1.12}(c)), whereas at the higher \( \mathrm{Re}_i = 18,900 \)  it undergoes terminal ascending motion following a primary pinch-off (see Fig.~\ref{fig:sphere 1.12}(d)). The corresponding dimensionless displaced volumes $V_{\text{Disp}}^{\text{exp}^*}$ are 1.25 and 1.34, respectively, indicating a larger entrained cavity volume in the latter case. This increased entrained cavity volume results in a larger net buoyancy force, causing the sphere to reverse its trajectory and exhibit terminally ascending motion. 

\begin{figure}
    \begin{subfigure}[t]{0.33\textwidth}
    \includegraphics[width=\linewidth]{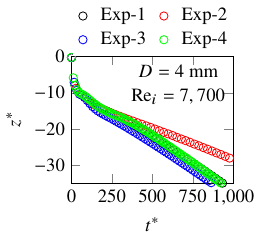}
 \caption{}
   \end{subfigure}
   \hfill
   \begin{subfigure}[t]{0.32\textwidth}
   \includegraphics[width=0.9\linewidth]{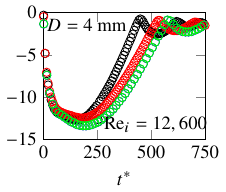}
    \caption{}
   \end{subfigure}
   \hfill
   \begin{subfigure}[t]{0.32\textwidth}
   \includegraphics[width=0.9\linewidth]{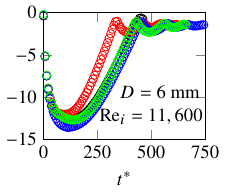}
    \caption{}
   \end{subfigure}
   \hfill
   \begin{subfigure}[t]{0.32\textwidth}
   \includegraphics[width=0.96\linewidth,trim=1mm 0 0 0, clip]{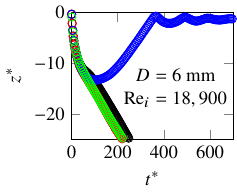}
    \caption{}
   \end{subfigure}
   \hfill
   \begin{subfigure}[t]{0.32\textwidth}
   \includegraphics[width=0.9\linewidth]{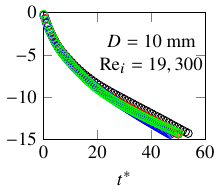}
 \caption{}
   \end{subfigure}
   \hfill
   \begin{subfigure}[t]{0.32\textwidth}
   \includegraphics[width=0.9\linewidth]{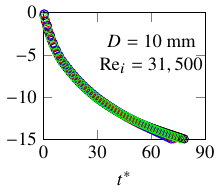}
 \caption{}
   \end{subfigure}
    \caption{(color online) Repeated experimental depth trajectories ($z^*$) of $\rho^* = 1.37$ sphere. (a) $D =$ 4~mm, Re$_i$~=~7,700; (b) $D =$ 4~mm, Re$_i$~=~12,600; (c) $D =$ 6~mm, Re$_i$~=~11,600; (d) $D =$ 6~mm, Re$_i$~=~18,900; (e) $D =$ 10~mm, Re$_i$~=~19,300; (f) $D =$ 10~mm, Re$_i$~=~31,500. The legend in (a) is applicable to (b)-(f).}
    \label{fig:sphere 1.37}
\end{figure}

For \( \rho^* = 1.37 \), Fig.~\ref{fig:sphere 1.37} illustrates the various trajectories observed across different impact Reynolds numbers. In contrast to the \( \rho^* = 1.12 \), \( D = 4\,\text{mm} \) case, the \( \rho^* = 1.37 \), \( D = 4\,\text{mm} \) spheres exhibit terminally descending motion at \( \mathrm{Re}_i = 7,700 \) (Fig.~\ref{fig:sphere 1.37}(a)), corresponding to $V_{\text{Disp}}^{\text{exp}^*} \approx$ 1.66. However, at a higher \( \mathrm{Re}_i = 12{,}600 \), as shown in Fig.~\ref{fig:sphere 1.37}(b), the sphere exhibits terminally ascending motion with $V_{\text{Disp}}^{\text{exp}^*} \approx$ 1.73,  again attributed to the strong buoyancy force due to larger cavity entrainment. At \( \mathrm{Re}_i = 11,600 \), the \( D = 6\,\text{mm} \) (see Fig.~\ref{fig:sphere 1.37}(c)) spheres exhibit terminally ascending motion with a relatively large displaced volume $V_{\text{Disp}}^{\text{exp}^*} \approx$ 1.89, resulting solely from a primary pinch-off. In contrast, at \( \mathrm{Re}_i = 18,900 \) (shown in Fig.~\ref{fig:sphere 1.37}(d)), although one experiment exhibits terminal ascending motion with a primary pinch-off with $V_{\text{Disp}}^{\text{exp}^*} \approx$  1.64, all other cases with $V_{\text{Disp}}^{\text{exp}^*}<1.64$ result in terminal descending motion with a secondary pinch-off. For the \( D = 10\,\text{mm} \) spheres, terminal descending motion is consistently identified across all $\mathrm{Re}_i$, with a secondary pinch-off, indicating that the buoyancy force generated is insufficient to overcome the downward forces acting on the sphere.

\begin{figure}
    \begin{subfigure}{0.34\textwidth}
    \includegraphics[width=\linewidth]{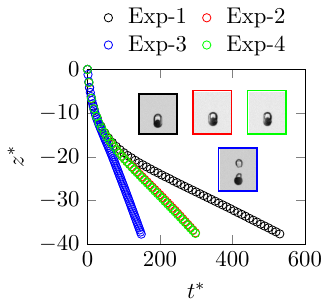}
 \caption{}
   \end{subfigure}
   \hfill
   \begin{subfigure}{0.31\textwidth}
   \includegraphics[width=\linewidth]{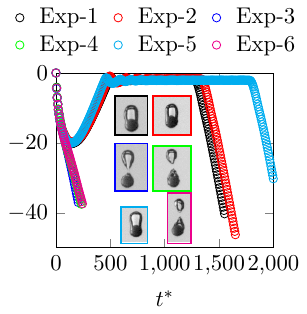}
 \caption{}
   \end{subfigure}
   \hfill
   \begin{subfigure}{0.31\textwidth}
   \includegraphics[width=\linewidth]{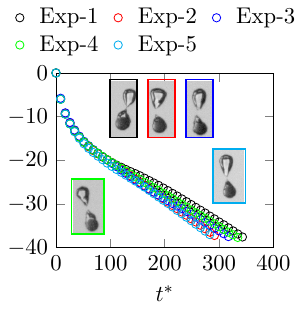}
 \caption{}
   \end{subfigure}
    \caption{(color online) Depth trajectory of a 4~mm sphere of dimensionless density $\rho^* = 2.16$ shown for repeated experiments and at three different impact Reynolds numbers: (a) 6,300; (b) 10,000; (c) 15,400.  Insets highlight the wake air cavity and the primary and secondary pinch-off events. }
    \label{fig:D4_H300_600}
\end{figure}

Less consistent are the trajectories for a sphere of density ratio $\rho^* = 2.16$, as shown in Fig.~\ref{fig:D4_H300_600} for a $D=4$~mm diameter sphere for three different impact Reynolds numbers and multiple repeated experiments. At both the lowest and highest Reynolds number, 6,300 and 15,400, the sphere exhibits secondary pinch-off and terminally descends (Fig.~\ref{fig:D4_H300_600}(a) and (c)). The descent velocity depends on the size of the remaining wake cavity, especially evident in Fig.~\ref{fig:D4_H300_600}(a). However, at the intermediate Reynolds number of 10,000, the sphere exhibits both terminally descending and ascending motion. The spheres which ascend back to the free surface are then eventually wetted to a higher degree, and descend with a smaller wake cavity. What these results indicate is, that the occurrence of a secondary pinch-off is not alone a function of the impact Reynolds number.

\begin{figure}[t]
\centering
  \begin{subfigure}[t]{0.48\textwidth}
    \centering
    \includegraphics[width=\textwidth]{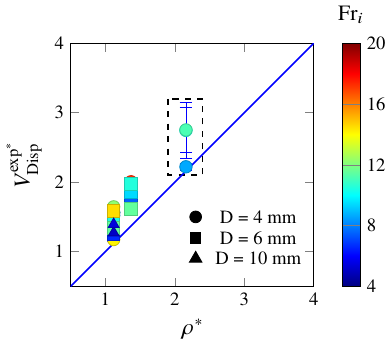}
    \caption{}
  \end{subfigure}
  \hfill
  \begin{subfigure}[t]{0.48\textwidth}
    \centering
    \includegraphics[width=\textwidth]{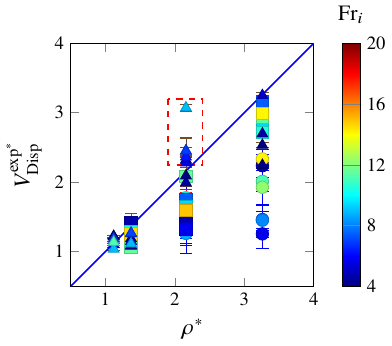}
    \caption{}
  \end{subfigure}
\caption{(color online) Dimensionless displaced volumes plotted against dimensionless density ratios. According to Eq.~(\ref{eq:ascending condition}), points above the diagonal are terminally ascending and below are terminally descending. (a) Values shown following only the primary pinch-off. The dashed  box indicates a regime in which spheres exhibit both terminally ascending and descending trajectories under the same impact conditions.(b) Values following the last pinch-off observed within the field of view.  The red dashed box marks a regime in which multiple pinch-off events are further observed while descending. The legend in (a) applies to (b).} 
\label{fig:Displaced_volumes}
\end{figure}

The results of all trials are summarized in Fig.~\ref{fig:Displaced_volumes}, plotting $V_{\text{Disp}}^{\text{exp}^*}$ against $\rho^*$. Figure~\ref{fig:Displaced_volumes}(a) summarizes those experiments in which the sphere is terminally ascending, while Fig.~\ref{fig:Displaced_volumes}(b) is for terminally descending spheres. Note that all spheres with dimensionless densities $\rho^*\ge 3.26$ are terminal descending; however, the $V_{\text{Disp}}^{\text{exp}^*}$ following the secondary pinch-off was out of the camera field of view. In Fig.~\ref{fig:Displaced_volumes}(a) the dashed box indicates conditions in which the sphere can either be terminally ascending or descending. 

Figure~\ref{fig:Displaced_volumes}(b) presents the $V_{\text{Disp}}^{\text{exp}^*}$ value after the last pinch-off within the observed field of view. In some cases, this corresponds to a secondary or multiple pinch-offs, and thus these spheres, being below the diagonal, are terminally descending. In the case of $\rho^* = 2.16$, larger diameter spheres (e.g. 10~mm) undergo additional fragmentation, resulting in tertiary pinch-off events as shown in the red dashed box in Fig.~\ref{fig:Displaced_volumes}(b). While these higher-order pinch-off events result in terminal descent, they cannot be captured within the field of view; hence, the $V_{\text{Disp}}^{\text{exp}^*}$ still appears above the diagonal.

\section{Conclusions}
\label{sec:conclusion}

This study reveals a number of novel aspects of a superhydrophobic sphere penetrating a deep pool, improving  the capability of predicting its trajectory over longer periods of time. This pertains especially to spheres of lighter density ratio, which hitherto have received little attention in the literature.

Although cavity pinch-off has been the subject of many earlier studies, its implication on the buoyancy force and the ensuing change in sphere deceleration has not previously been exposed. Similarly, the pinch-off event results in an additional, albeit short-lived downward force acting on the sphere, making itself apparent as reduced drag.  Furthermore, the size of the resulting gas cavity attached to the penetrating sphere scales very closely with the impact Froude number and density ratio. The importance of secondary or further pinch-off events has also been underlined, often determining whether the sphere will continue descending in the pool or will ascend after reaching a minimum depth and zero velocity. Terminally ascending is a trajectory only observed for lighter spheres, as expected.

Despite these new insights into the penetration of superhydrophobic spheres into a deep pool, numerous questions remain for future investigation and discussion. Above all, the forces associated with virtual mass and the history term in the BBO equation remain unresolved during the initial period of strong sphere deceleration. We have lumped these terms into the hydrodynamic force, which also includes viscous and pressure drag; however, also drag is expected to deviate strongly between values for steady flow and values during strong deceleration. Indeed it is questionable whether it is even appropriate (or possible) to differentiate among these three forces during deceleration, since their deviations from steady flow values are all rooted in the same time response of the boundary layer to a change in the outer flow. 

The superhydrophobic sphere represents a middle ground between hydrophilic \cite{Billa_Josyula_Tropea_Mahapatra_2025} and hydrophobic spheres which remain completely unwetted, e.g. a Leidenfrost sphere \cite{jetly2018drag,Vakarelski2014LeidenfrostWater}. Fig.~\ref{fig:Result_NC10mm_Re15700} already showed a large variation in the trajectory behavior between hydrophilic and superhydrophobic, attributed to the degree of wake symmetry. The question therefore arises: how does the wake symmetry depend on the degree of hydrophobicity?
Experimentally, this influence could only be systematically investigated with a three-dimensional and temporal resolution of the contact line on the sphere during its descent - not an easy experiment!

\section*{Appendix}

\begin{figure}
    \begin{subfigure}{0.44\textwidth}
        \includegraphics[width=\textwidth]{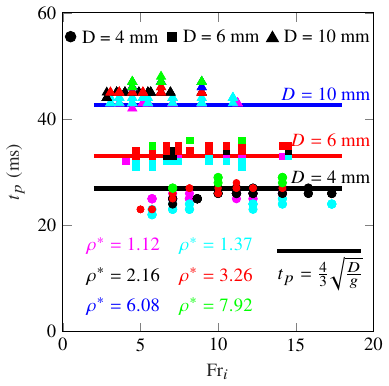}
        \caption{}
    \end{subfigure}
    \hspace{0.02\textwidth}
    \begin{subfigure}{0.45\textwidth}
        \includegraphics[width=\textwidth]{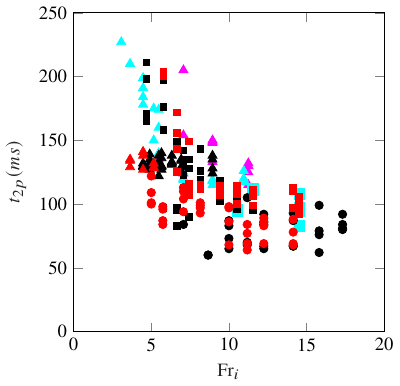}
        \caption{}
    \end{subfigure}
    \caption{(color online) Dimensional pinch-off times as a function of $\mathrm{Fr}_i$. (a) dimensional primary pinch-off time ($t_p$) and (b) dimensional secondary pinch-off time ($t_{2p}$). The solid colored bars correspond to the pinch-off time predicted using Eq.~(\ref{eq:t_Fr}). The legends and colors in (a) also apply to (b). }
\label{fig:Dimensional_pinch_off_time}
\end{figure}

The dimensional primary pinch-off time $t_{p}$, shown in Fig.~\ref{fig:Dimensional_pinch_off_time}(a), remains nearly invariant with the impact Froude number $\mathrm{Fr}_i$ across all investigated diameters, in agreement with the scaling given in Eq.~(\ref{eq:t_Fr})
The result indicates that the primary pinch-off is governed predominantly by hydrostatic pressure and scales with the diameter of the sphere and gravity. 

In contrast, the secondary pinch-off time $t_{2p}$ in Fig.~\ref{fig:Dimensional_pinch_off_time}(b) shows a strong dependence on $\mathrm{Fr}_i$, decreasing with increasing impact velocity ($v_i$). Longer (and deeper) cavities formed at higher velocity collapse faster under stronger hydrostatic pressure, leading to shorter $t_{2p}$. The scatter across density ratios reflects the influence of sphere inertia: lighter spheres tend to produce slightly delayed secondary pinch-offs, whereas denser spheres close more rapidly. Thus, while the primary pinch-off is governed mainly by diameter and gravity, the secondary pinch-off reflects a dynamic balance between cavity length, density of the sphere, and hydrostatic pressure recovery.

\begin{acknowledgments}
CT acknowledges support from the Indian Institute of Technology Madras through the appointment as the Henry Ford Chair Professorship. PSM acknowledges the V. Ganesan Faculty Fellowship received from IIT Madras. The authors would like to express their sincere gratitude to Dr. Tejaswi Josyula, whose immense efforts in the experimental set-up, meticulous image processing, plotting support, and insightful discussions were invaluable to this work. The authors would like to thank Dr. Arvind Pattamatta for providing access to the high-speed camera.
\end{acknowledgments}

\section*{Declaration of interests}
The authors report no potential conflicts of interest.

\bibliography{references.bib}

\begin{thebibliography}{46}
\providecommand{\natexlab}[1]{#1}
\providecommand{\url}[1]{\texttt{#1}}
\expandafter\ifx\csname urlstyle\endcsname\relax
  \providecommand{\doi}[1]{doi: #1}\else
  \providecommand{\doi}{doi: \begingroup \urlstyle{rm}\Url}\fi

\bibitem[Aristoff and Bush(2009)]{Aristoff2009WaterSpheres}
Jeffrey~M. Aristoff and John~W.M. Bush.
\newblock {Water entry of small hydrophobic spheres}.
\newblock \emph{Journal of Fluid Mechanics}, 619:\penalty0 45--78, 2009.

\bibitem[Aristoff et~al.(2008)Aristoff, Truscott, Techet, and
  Bush]{aristoff2008water}
Jeffrey~M Aristoff, Tadd~T Truscott, Alexandra~H Techet, and John~WM Bush.
\newblock The water-entry cavity formed by low {B}ond number impacts.
\newblock \emph{Physics of Fluids}, 20\penalty0 (9), 2008.

\bibitem[Aristoff et~al.(2010)Aristoff, Truscott, Techet, and
  Bush]{Aristoff2010TheSpheres}
Jeffrey~M. Aristoff, Tadd~T. Truscott, Alexandra~H. Techet, and John~W.M. Bush.
\newblock {The water entry of decelerating spheres}.
\newblock \emph{Physics of Fluids}, 22\penalty0 (3):\penalty0 1--8, 2010.

\bibitem[Asfar and Moore(1987)]{asfar1987rigid}
K~Asfar and S~Moore.
\newblock Rigid-body water impact at shallow angles of incidence.
\newblock In \emph{Proceedings of the Sixth International Offshore Mechanics
  and Arctic Engineering Symposium}, pages 105--112, 1987.

\bibitem[Billa et~al.(2025)Billa, Josyula, Tropea, and
  Mahapatra]{Billa_Josyula_Tropea_Mahapatra_2025}
Prasanna~Kumar Billa, Tejaswi Josyula, Cameron Tropea, and Pallab~Sinha
  Mahapatra.
\newblock Motion of a rigid sphere penetrating a deep pool.
\newblock \emph{Journal of Fluid Mechanics}, 1012:\penalty0 A5, 2025.

\bibitem[Brown and Lawler(2003)]{Brown2003SphereRevisited}
Phillip~P. Brown and Desmond~F. Lawler.
\newblock {Sphere drag and settling velocity revisited}.
\newblock \emph{Journal of Environmental Engineering}, 129\penalty0
  (3):\penalty0 222--231, 2003.

\bibitem[Crowe et~al.(2011)Crowe, Schwarzkopf, Sommerfeld, and
  Tsuji]{crowe2011multiphase}
Clayton~T Crowe, John~D Schwarzkopf, Martin Sommerfeld, and Yutaka Tsuji.
\newblock \emph{Multiphase {F}lows with {D}roplets and {P}articles}.
\newblock CRC Press LLC Boca Raton, FL, 2011.

\bibitem[Dong(1978)]{dong1978effective}
RG~Dong.
\newblock Effective mass and damping of submerged structures.
\newblock Technical report, Lawrence Livermore National Lab.(LLNL), Livermore,
  CA (United States), 1978.

\bibitem[Duclaux et~al.(2007)Duclaux, Caill{\'{e}}, Duez, Ybert, Bocquet, and
  Clanet]{Duclaux2007DynamicsCavities}
V.~Duclaux, F.~Caill{\'{e}}, C.~Duez, C.~Ybert, L.~Bocquet, and C.~Clanet.
\newblock {Dynamics of transient cavities}.
\newblock \emph{Journal of Fluid Mechanics}, 591:\penalty0 1--19, 2007.

\bibitem[Duez et~al.(2007)Duez, Ybert, Clanet, and Bocquet]{duez2007making}
Cyril Duez, Christophe Ybert, Christophe Clanet, and Lyderic Bocquet.
\newblock Making a splash with water repellency.
\newblock \emph{Nature Physics}, 3\penalty0 (3):\penalty0 180--183, 2007.

\bibitem[Extrand(2002)]{Extrand2002WaterSurfaces}
C.~W. Extrand.
\newblock {Water contact angles and hysteresis of polyamide surfaces}.
\newblock \emph{Journal of Colloid and Interface Science}, 248\penalty0
  (1):\penalty0 136--142, 2002.

\bibitem[Extrand and Moon(2012)]{Extrand2012IndirectSurfaces}
C.~W. Extrand and Sung~In Moon.
\newblock {Indirect methods to measure wetting and contact angles on spherical
  convex and concave surfaces}.
\newblock \emph{Langmuir}, 28\penalty0 (20):\penalty0 7775--7779, 2012.

\bibitem[Guleria et~al.(2021)Guleria, Dhar, and
  Patil]{Guleria2021ExperimentalSphere}
Sharey~Deep Guleria, Atul Dhar, and Dhiraj~V. Patil.
\newblock {Experimental insights on the water entry of hydrophobic sphere}.
\newblock \emph{Physics of Fluids}, 33\penalty0 (10), 2021.

\bibitem[Gupta et~al.(2016)Gupta, Vaikuntanathan, and
  Sivakumar]{gupta2016superhydrophobic}
R~Gupta, V~Vaikuntanathan, and D~Sivakumar.
\newblock Superhydrophobic qualities of an aluminum surface coated with
  hydrophobic solution neverwet.
\newblock \emph{Colloids and Surfaces A: Physicochemical and Engineering
  Aspects}, 500:\penalty0 45--53, 2016.

\bibitem[Horowitz and Williamson(2008)]{horowitz2008critical}
M~Horowitz and CHK Williamson.
\newblock Critical mass and a new periodic four-ring vortex wake mode for
  freely rising and falling spheres.
\newblock \emph{Physics of Fluids}, 20\penalty0 (10), 2008.

\bibitem[Horowitz and Williamson(2010)]{horowitz2010effect}
M~Horowitz and CHK Williamson.
\newblock The effect of {R}eynolds number on the dynamics and wakes of freely
  rising and falling spheres.
\newblock \emph{Journal of Fluid Mechanics}, 651:\penalty0 251--294, 2010.

\bibitem[Jetly et~al.(2018)Jetly, Vakarelski, and Thoroddsen]{jetly2018drag}
Aditya Jetly, Ivan~U Vakarelski, and Sigurdur~T Thoroddsen.
\newblock Drag crisis moderation by thin air layers sustained on
  superhydrophobic spheres falling in water.
\newblock \emph{Soft Matter}, 14\penalty0 (9):\penalty0 1608--1613, 2018.

\bibitem[Kuwabara et~al.(1983)Kuwabara, Chiba, and Kono]{kuwabara1983anomalous}
Goro Kuwabara, Seiji Chiba, and Kimitoshi Kono.
\newblock Anomalous motion of a sphere falling through water.
\newblock \emph{Journal of the Physical Society of Japan}, 52\penalty0
  (10):\penalty0 3373--3381, 1983.

\bibitem[Li et~al.(2019)Li, Zhang, Zhang, Huang, Ma, and
  Wang]{li2019experimental}
Daqin Li, Jiayue Zhang, Mindi Zhang, Biao Huang, Xiaojian Ma, and Guoyu Wang.
\newblock Experimental study on water entry of spheres with different surface
  wettability.
\newblock \emph{Ocean Engineering}, 187:\penalty0 106123, 2019.

\bibitem[Magos and Balan(2021)]{Magos2021ContactSurfaces}
Istvan Magos and Corneliu Balan.
\newblock {Contact Angles on Spherical Hydrophilic Surfaces}.
\newblock \emph{12th International Symposium on Advanced Topics in Electrical
  Engineering, ATEE 2021}, 21\penalty0 (9):\penalty0 9470--9473, 2021.

\bibitem[Mansoor et~al.(2014)Mansoor, Marston, Vakarelski, and
  Thoroddsen]{Mansoor2014WaterFormation}
M.~M. Mansoor, J.~O. Marston, I.~U. Vakarelski, and S.~T. Thoroddsen.
\newblock {Water entry without surface seal: Extended cavity formation}.
\newblock \emph{Journal of Fluid Mechanics}, 743:\penalty0 295--326, 2014.

\bibitem[Mansoor et~al.(2017)Mansoor, Vakarelski, Marston, Truscott, and
  Thoroddsen]{Mansoor2017Stable-streamlinedSpheres}
M.~M. Mansoor, I.~U. Vakarelski, J.~O. Marston, T.~T. Truscott, and S.~T.
  Thoroddsen.
\newblock {Stable-streamlined and helical cavities following the impact of
  Leidenfrost spheres}.
\newblock \emph{Journal of Fluid Mechanics}, 823:\penalty0 716--754, 2017.

\bibitem[Marston et~al.(2016)Marston, Truscott, Speirs, Mansoor, and
  Thoroddsen]{marston2016crown}
Jeremy~O Marston, Tadd~T Truscott, Nathan~B Speirs, Mohammad~M Mansoor, and
  Sigurdur~T Thoroddsen.
\newblock Crown sealing and buckling instability during water entry of spheres.
\newblock \emph{Journal of Fluid Mechanics}, 794:\penalty0 506--529, 2016.

\bibitem[May(1951)]{may1951effect}
Albert May.
\newblock Effect of surface condition of a sphere on its water-entry cavity.
\newblock \emph{Journal of Applied Physics}, 22\penalty0 (10):\penalty0
  1219--1222, 1951.

\bibitem[McHale et~al.(2009)McHale, Shirtcliffe, Evans, and
  Newton]{mchale2009terminal}
Glen McHale, NJ~Shirtcliffe, CR~Evans, and MI~Newton.
\newblock Terminal velocity and drag reduction measurements on superhydrophobic
  spheres.
\newblock \emph{Applied Physics Letters}, 94\penalty0 (6):\penalty0 064104,
  2009.

\bibitem[Pallas and Pethica(1983)]{pallas1983surface}
NR~Pallas and BA~Pethica.
\newblock The surface tension of water.
\newblock \emph{Colloids and Surfaces}, 6\penalty0 (3):\penalty0 221--227,
  1983.

\bibitem[Plesset and Prosperetti(1977)]{plesset1977bubble}
Milton~S Plesset and Andrea Prosperetti.
\newblock Bubble dynamics and cavitation.
\newblock \emph{Annual Review of Fluid Mechanics}, 9\penalty0 (1):\penalty0
  145--185, 1977.

\bibitem[Rayleigh(1917)]{rayleigh1917viii}
Lord Rayleigh.
\newblock {VIII}. {O}n the pressure developed in a liquid during the collapse
  of a spherical cavity.
\newblock \emph{The London, Edinburgh, and Dublin Philosophical Magazine and
  Journal of Science}, 34\penalty0 (200):\penalty0 94--98, 1917.

\bibitem[Scoggins(1967)]{scoggins1967sphere}
James~R Scoggins.
\newblock \emph{Sphere behavior and the measurement of wind profiles}.
\newblock National Aeronautics and Space Administration, 1967.

\bibitem[Speirs et~al.(2019{\natexlab{a}})Speirs, Belden, Pan, Holekamp,
  Badlissi, Jones, and Truscott]{Speirs2019TheJet}
Nathan~B. Speirs, Jesse Belden, Zhao Pan, Sean Holekamp, George Badlissi,
  Matthew Jones, and Tadd~T. Truscott.
\newblock {The water entry of a sphere in a jet}.
\newblock \emph{Journal of Fluid Mechanics}, 863:\penalty0 956--968,
  2019{\natexlab{a}}.

\bibitem[Speirs et~al.(2019{\natexlab{b}})Speirs, Mansoor, Belden, and
  Truscott]{Speirs2019WaterAnglesb}
Nathan~B. Speirs, Mohammad~M. Mansoor, Jesse Belden, and Tadd~T. Truscott.
\newblock {Water entry of spheres with various contact angles}.
\newblock \emph{Journal of Fluid Mechanics}, 862:\penalty0 1--13,
  2019{\natexlab{b}}.

\bibitem[Spurk and Aksel(2007 (Eq. (5.45)))]{spurk2007fluid}
Joseph Spurk and Nuri Aksel.
\newblock \emph{Fluid Mechanics}.
\newblock Springer Science \& Business Media, 2007 (Eq. (5.45)).

\bibitem[Sun et~al.(2019)Sun, Wang, Zong, Zhang, Wang, and Xu]{sun2019splash}
Tiezhi Sun, Heng Wang, Zhi Zong, Guiyong Zhang, An~Wang, and Chang Xu.
\newblock Splash formation and cavity dynamics of sphere entry through a
  viscous liquid resting on the water.
\newblock \emph{{AIP} Advances}, 9\penalty0 (7), 2019.

\bibitem[Taneda(1978)]{taneda1978visual}
Sadatoshi Taneda.
\newblock Visual observations of the flow past a sphere at reynolds numbers
  between 104 and 106.
\newblock \emph{Journal of Fluid Mechanics}, 85\penalty0 (1):\penalty0
  187--192, 1978.

\bibitem[Techet and Truscott(2011)]{techet2011water}
AH~Techet and TT~Truscott.
\newblock Water entry of spinning hydrophobic and hydrophilic spheres.
\newblock \emph{Journal of Fluids and Structures}, 27\penalty0 (5-6):\penalty0
  716--726, 2011.

\bibitem[Truscott et~al.(2012)Truscott, Epps, and
  Techet]{Truscott2012UnsteadyEntry}
Tadd~T. Truscott, Brenden~P. Epps, and Alexandra~H. Techet.
\newblock {Unsteady forces on spheres during free-surface water entry}.
\newblock \emph{Journal of Fluid Mechanics}, 704:\penalty0 173--210, 2012.

\bibitem[Truscott et~al.(2014)Truscott, Epps, and
  Belden]{Truscott2014WaterProjectiles}
Tadd~T. Truscott, Brenden~P. Epps, and Jesse Belden.
\newblock {Water entry of projectiles}.
\newblock \emph{Annual Review of Fluid Mechanics}, 46:\penalty0 355--378, 2014.

\bibitem[Truscott(2009)]{truscott2009cavity}
Tadd~Trevor Truscott.
\newblock \emph{Cavity dynamics of water entry for spheres and ballistic
  projectiles}.
\newblock Ph.d. thesis, Massachusetts Institute of Technology, Cambridge, MA,
  June 2009.

\bibitem[Vakarelski et~al.(2011)Vakarelski, Marston, Chan, and
  Thoroddsen]{vakarelski2011drag}
Ivan~U Vakarelski, Jeremy~O Marston, Derek~YC Chan, and Sigurdur~T Thoroddsen.
\newblock Drag reduction by {L}eidenfrost vapor layers.
\newblock \emph{Physical Review Letters}, 106\penalty0 (21):\penalty0 214501,
  2011.

\bibitem[Vakarelski et~al.(2014)Vakarelski, Chan, and
  Thoroddsen]{Vakarelski2014LeidenfrostWater}
Ivan~U. Vakarelski, Derek~Y.C. Chan, and Sigurdur~T. Thoroddsen.
\newblock {Leidenfrost vapour layer moderation of the drag crisis and
  trajectories of superhydrophobic and hydrophilic spheres falling in water}.
\newblock \emph{Soft Matter}, 10\penalty0 (31):\penalty0 5662--5668, 2014.

\bibitem[Vakarelski et~al.(2017)Vakarelski, Klaseboer, Jetly, Mansoor,
  Aguirre-Pablo, Chan, and Thoroddsen]{Vakarelski2017Self-determinedCavities}
Ivan~U. Vakarelski, Evert Klaseboer, Aditya Jetly, Mohammad~M. Mansoor,
  Andres~A. Aguirre-Pablo, Derek~Y.C. Chan, and Sigurdur~T. Thoroddsen.
\newblock {Self-determined shapes and velocities of giant near-zero drag gas
  cavities}.
\newblock \emph{Science Advances}, 3\penalty0 (9):\penalty0 1--8, 2017.

\bibitem[Veldhuis and Biesheuvel(2007)]{veldhuis2007experimental}
CHJ Veldhuis and A~Biesheuvel.
\newblock An experimental study of the regimes of motion of spheres falling or
  ascending freely in a {N}ewtonian fluid.
\newblock \emph{International Journal of Multiphase Flow}, 33\penalty0
  (10):\penalty0 1074--1087, 2007.

\bibitem[Veldhuis et~al.(2005)Veldhuis, Biesheuvel, Van~Wijngaarden, and
  Lohse]{Veldhuis2005MotionParticles}
Christian Veldhuis, Arie Biesheuvel, Leen Van~Wijngaarden, and Detlef Lohse.
\newblock {Motion and wake structure of spherical particles}.
\newblock \emph{Nonlinearity}, 18\penalty0 (1), 2005.

\bibitem[Watson et~al.(2020)Watson, Souchik, Weinberg, Bom, and
  Dickerson]{watson2020making}
Daren~A Watson, Chris~J Souchik, Madison~P Weinberg, Joshua~M Bom, and Andrew~K
  Dickerson.
\newblock Making a splash with fabrics in hydrophilic sphere entry.
\newblock \emph{Journal of Fluids and Structures}, 94:\penalty0 102907, 2020.

\bibitem[Weisensee et~al.(2016)Weisensee, Tian, Miljkovic, and
  King]{weisensee2016water}
Patricia~B Weisensee, Junjiao Tian, Nenad Miljkovic, and William~P King.
\newblock Water droplet impact on elastic superhydrophobic surfaces.
\newblock \emph{Scientific Reports}, 6\penalty0 (1):\penalty0 30328, 2016.

\bibitem[Zhu and Fan(1998)]{zhu1998multiphase}
C~Zhu and LS~Fan.
\newblock Multiphase flow: Gas/solid.
\newblock In \emph{The Handbook of Fluid Dynamics}. CRC Press LLC Boca Raton,
  FL, 1998.

\end{thebibliography}

\end{document}